\documentclass[apjl]{emulateapj}

\usepackage{graphicx}
\usepackage{t1enc}
\usepackage[varg]{txfonts}
\usepackage{epsfig}
\usepackage{rotating}
\usepackage{natbib}
\usepackage{color}

\definecolor{RED}{rgb}{1.0,0.0,0.0}
\def\aap{A\& A}

\def\apjs{ApJS}
\def\apjl{ApJL}

\shorttitle{Converging flows in NGC 6334 V}
\shortauthors{Ju\'arez et al.}

\begin{document}

\title{Magnetized converging flows towards the hot core in the intermediate/high-mass star-forming region NGC 6334 V}

\author{Carmen Ju\'arez\altaffilmark{1,2},
Josep M. Girart\altaffilmark{1},
Manuel Zamora-Avil\'es\altaffilmark{3,4},
Ya-Wen Tang\altaffilmark{5},
Patrick M. Koch\altaffilmark{5},
Hauyu Baobab Liu\altaffilmark{6},\\
Aina Palau\altaffilmark{3},
Javier Ballesteros-Paredes\altaffilmark{3},
Qizhou Zhang\altaffilmark{7},
Keping Qiu\altaffilmark{8}
}
\altaffiltext{1}{Institut de Ci\`encies de l'Espai, (CSIC-IEEC), Campus UAB, Carrer de Can Magrans, S/N, 08193 Cerdanyola del Vall\`es, Catalonia, Spain}
\email{juarez@ice.cat}
\altaffiltext{2}{Dpt. de F\'isica Qu\`antica i Astrof\'isica, Institut de Ci\`encies del Cosmos (ICCUB), Universitat de Barcelona (IEEC-UB), Mart\'i i Franqu\`es 1, 08028 Barcelona, Catalonia, Spain} 
\altaffiltext{3}{Instituto de Radioastronom\'ia y Astrof\'isica, Universidad Nacional Aut\'onoma de M\'exico, P.O. Box 3-72, 58090, Morelia, Michoac\'an, Mexico}
\altaffiltext{4}{Department of Astronomy, University of Michigan, 311 West Hall, 1085 S. University Ann Arbor, MI 48109-1107, USA}
\altaffiltext{5}{Academia Sinica Institute of Astronomy and Astrophysics, P.O. Box 23-141, Taipei, 10617 Taiwan}
\altaffiltext{6}{European Southern Observatory (ESO), Karl-Schwarzschild-Str. 2, D-85748 Garching, Germany}
\altaffiltext{7}{Harvard-Smithsonian Center for Astrophysics, 60 Garden Street, Cambridge, MA 02138, USA}
\altaffiltext{8}{School of Astronomy and Space Science, Nanjing University, 22 Hankou Road, Nanjing, Jiangsu 210093, China}

\begin{abstract}
We present Submillimeter Array (SMA) observations at 345 GHz towards the intermediate/high-mass cluster-forming region NGC 6334 V. From the dust emission we spatially resolve three dense condensations, the brightest one presenting the typical chemistry of a hot core. The magnetic field (derived from the dust polarized emission) shows a bimodal converging pattern towards the hot core. The molecular emission traces two filamentary structures at two different velocities, separated by 2 km s$^{-1}$, converging to the hot core and following the magnetic field distribution. We compare the velocity field and the magnetic field derived from the SMA observations with MHD simulations of star-forming regions dominated by gravity. This comparison allows us to show how the gas falls in from the larger-scale extended dense core ($\sim0.1$~pc) of NGC 6334 V towards the higher-density hot core region ($\sim0.02$~pc) through two distinctive converging flows dragging the magnetic field, whose strength seems to have been overcome by gravity. 

\par
\end{abstract}


\section{Introduction}
Previous observations have led to the consensus that the magnetic (B) field strength is non-negligible during the formation of molecular gas clouds and dense gas filaments, for low- and high-mass star- and cluster-forming regions (e.g., \citealt{Matthews09}; \citealt{Crutcher10}; \citealt{Koch14}; \citealt{Fissel16}; \citealt{Soler16}; see \citealt{Li14} for a review).
However, the role of the magnetic field in the formation and evolution of dense molecular cores (which have sizes of $\sim0.1$~pc) is still a matter of debate \citep[e.g.,][]{Vazquez-Semadeni11,Bertram12,Crutcher12,Li14,Kortgen&Banerjee15}. Its role can be studied observationally by combining high angular resolution ($\sim$few arcseconds) observations of molecular lines with polarimetric observations of dust continuum \citep[e.g.,][]{Girart13,Hull14,Zhang14}. However, this type of studies has been done only towards a limited sample of star-forming cores. Within this sample, there are cases where the magnetic field is consistent with an hourglass morphology, which is an indication of strong magnetic fields in infalling cores \citep{Fiedler93,GalliandShu93a,GalliandShu93b,NakamuraandLi05,Frau11}. Examples include the low- and high-mass star-forming regions NGC1333 IRAS4A \citep{Girart06,Frau11,Liu16}, G31.41+0.31 \citep{Girart09}, W51e2 \citep{Tang09b}, W51 North \citep{Tang13}, G35.2-0.74N \citep{Qiu13}, G240.31+0.07 \citep{Qiu14}, W43-MM1 \citep{Cortes16}. 

NGC 6334 is one of the nearest high-mass star-forming regions, located at a distance of 1.3 kpc \citep{Chibueze14,Reid14}.
The dense molecular gas is dominated by a $\sim$20 pc long filamentary structure oriented along northeast-southwest direction. Along the filament there are several parsec-scale sites of active star formation \citep[e.g.,][and references therein]{Li15}.
Previous optical and submillimeter polarimetric observations found a coherent magnetic field from cloud to core scales with its main direction approximately perpendicular to the long axis of the dense filament. This led \citet{Li15} to suggest that NGC 6334 undergoes self-similar fragmentation regulated by the magnetic field. 

NGC 6334 V is the southernmost active star-forming site along the large-scale northeast-southwest dense gas filament of NGC 6334.
This region was not covered by the multi-scale studies of \citet{Li15}.
NGC 6334 V is located at the intersection between the main filament and a smaller filament that extends along the northwest-southeast direction \citep[e.g.,][]{Andre16}. This characteristic makes the NGC~6334~V star-forming core different from other cores along the main filament. Therefore, the study of this core, at scales similar to \citet{Li15}, can be used to test whether the physical properties of NGC~6334~V are affected by the convergence of these two filaments.

Signposts of recent and ongoing star formation in NGC~6334~V include its high far-infrared luminosity (L$_{\text{bol}}\sim10^4$ L$_{\odot}$), a CO molecular outflow \citep{Fischer82,Kraemer95,Zhang14} and H$_2$O and OH masers \citep{Forster89,Raimond69,Brooks01}. 
It also harbors several infrared sources \citep{Simon85,Harvey83,Harvey84,Kraemer99} and three faint associated radio sources \citep{Rengarajan96}. 

Previous studies \citep{Hashimoto07,Simpson09} suggest the existence of two independent outflows in NGC 6334 V, one oriented east-west along the line of sight and the other with a north-south orientation almost in the plane of the sky. These two outflows could be powered by the infrared sources KDJ3 and KDJ4 \citep{Kraemer99}.

In this paper, we present Submillimeter Array (SMA)\footnote{The Submillimeter Array is a joint project between the Smithsonian Astrophysical Observatory and the Academia Sinica Institute of Astronomy and Astrophysics \citep{Ho04}.} observations towards NGC 6334 V at 345 GHz. These observations covered a frequency range that allows us to detect the dust emission with a high signal-to-noise ratio together with linear polarization from the dust emission. Within the frequency range there are various molecular lines (see section below for more details on the specific lines): dense gas tracers (i. e., sensitive to densities of $\gtrsim10^6$~cm$^{-3}$ and temperature $\gtrsim100$~K) and tracers of outflows or shocks. This allows us to characterize the physical properties of the core, and in particular to shed light on the role of the magnetic field.

The details of our observations are given in Section \ref{sec:obs}.
The observational results are shown in Section \ref{sec:results}.
In Section \ref{sec:analysis} we present magneto-hydrodynamic (MHD) simulations. Radiative transfer analysis and convolution with the SMA response are performed to properly compare with the observations. Discussion and conclusions are given in Section~\ref{sec:discussion} and \ref{sec:conclusion}, respectively.


\section{SMA observations and data reduction}
\label{sec:obs}
NGC 6334 V was observed with the SMA in the subcompact, compact and extended configurations. These observations were part of the polarization legacy project carried out between 2008 and 2012 in the 345 GHz band to observe the polarization emission of a sample of 21 massive star-forming regions \citep[for an overview of this survey see][]{Zhang14}.
Table~\ref{observations} shows the basic observational parameters and lists the calibrators used during the three tracks. The combined three configurations sample visibilities between $\sim$6 k$\lambda$ to $\sim$200 k$\lambda$. This makes the observations sensitive to scales between $1''$ and $15''$.
The phase and pointing center of the observations was R.A.=17:19:57.40, DEC=$-$35:57:46.0 (J2000). The frequency was approximately centered on the CO (3$-$2) line (345.79599 GHz) in the upper sideband (USB). The correlator consisted of 48 chunks with a bandwidth of 104 MHz each. This provided a total bandwidth (including the USB and the lower side band - LSB) of $\sim8$~GHz. We set the correlator to yield a uniform spectral resolution of 128 channels per chunk, providing a channel width of 0.8 MHz or 0.7~km~s$^{-1}$ at the observing frequency. 

Polarization observations were carried out using quarter-wave plates which convert linear polarization signals to circular ones \citep[see][for a more detailed description of the SMA polarimeter and the calibration process]{Marrone06,Marrone08}. The polarization leakages were measured to an accuracy of 0.1\% \citep{Marrone08}. The calibration for absolute flux, passband, and gains were carried out using the {\sc mir idl} software package \citep{Scoville93}. The images were created using the {\sc Miriad} software package \citep[][]{Sault95}. We averaged the line-free channels in the LSB and USB to generate the 870~$\mu$m continuum. The 870~$\mu$m maps were obtained combining the three configurations. We set the {\sc Robust} visibility weighting parameter to 0.5 \citep{Briggs95}, which provides a good compromise between angular resolution and sensitivity. The rms noise level of the resulting continuum map (Fig.~\ref{fig:continuum}) is 5.5 mJy beam$^{-1}$ and the synthesized beam is $2\farcs15\times1\farcs49$, P.A. $=-2.92^\circ$. To generate the polarization maps we used the {\sc impol} task in {\sc Miriad} which computes the linearly bias-corrected polarized intensity and position angle from Stokes Q and U (QU) images. We used 3$\sigma$ as the cutoff for Stokes QU \citep{Vaillancourt06} and I images. The rms level for Stokes QU is 3.1 mJy beam$^{-1}$. The different molecular line maps were also generated using a {\sc Robust} parameter of 0.5, except for the SiO (8$-$7) line. This line is strong and compact. Thus, we used a {\sc Robust} parameter of $-2$ (uniform weighting), which provides the best possible angular resolution.

\begin{table*}
\caption{SMA observations summary}
\begin{center}
\begin{tabular}{l c c c}
\hline
\hline
Array configuration		&subcompact			&compact			&extended			\\
\hline
Observing date			&2011 July 2 		&2011 June 18		&2011 July 21		\\
$\tau_{225\text{GHz}}$ 	&0.09				&0.06--0.1			&0.05				\\
Number of antennas		&7					&7					&8					\\
Flux calibrator			&Callisto			&Uranus		 		&Callisto			\\
Passband calibrator		&3C 279/3C 454.3 	&3C 279/3C 454.3	&3C 279/3C 454.3	\\
Gain calibrator			&1733-130			&1733-130 			&1733-130			\\
Polarization calibrator	&3C 279/3C 454.3	&3C 279/3C 454.3	&3C 279/3C 454.3	\\
uv range ($k\lambda$) 	&6--42				&8--80				&24--206 			\\
\hline	
\end{tabular}
\end{center}
\label{observations}
\end{table*}
  
\section{Results}	\label{sec:results}
\subsection{Morphology}
\subsubsection{Dust continuum} \label{continuum}
\begin{figure}
 \centering
\includegraphics[scale=0.45,keepaspectratio=true]{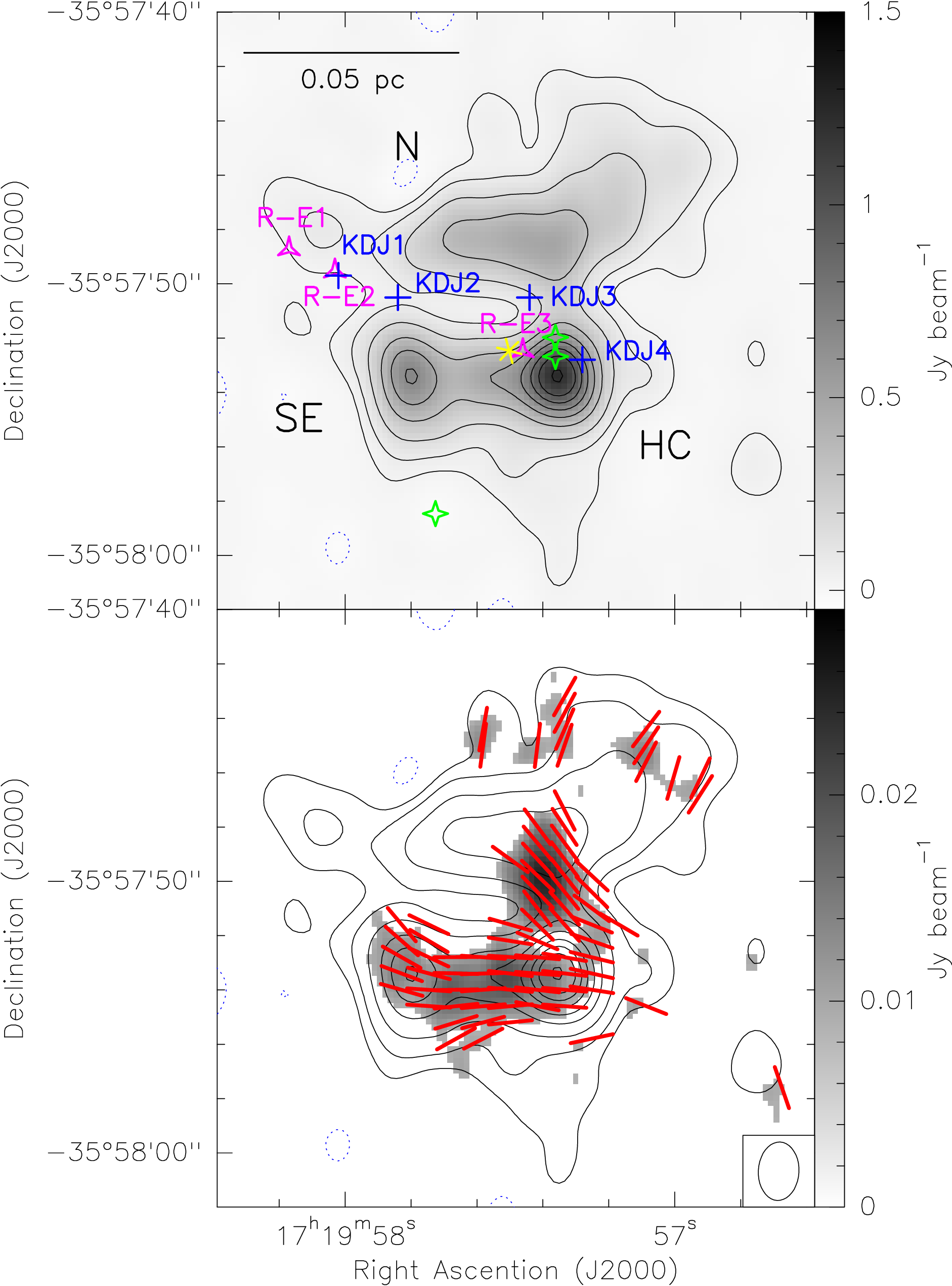}
  \caption{Images of 870 $\mu$m continuum observations on NGC 6334 V. {\it Upper panel}: The grey image and the contours show the 870 $\mu$m dust continuum emission. Contour levels are $-4$, 4, 12, 30, 60, ..., 180, 220 times the rms noise level of the map, 5.5 mJy beam$^{-1}$. The three main peaks are labeled as N (north), SE (southeast) and HC (hot core). Blue crosses, pink triangles, green stars and yellow asterisk show the positions of infrared \citep{Kraemer99}, radio \citep{Rengarajan96}, OH maser \citep{Raimond69,Brooks01} and H$_2$O maser sources \citep{Forster89}, respectively. The intensity scale bar (in Jy beam$^{-1}$) is located at the right side. A scale bar is located at the upper left corner. {\it Lower panel}: Grey scale image of the 870 $\mu$m dust polarization emission overlaid with the contour map of the 870 $\mu$m dust continuum emission. Grey scale shows the polarization intensity in Jy beam$^{-1}$. Red segments are the inferred magnetic field orientations projected on the plane of the sky (i.e. B-segments) using a Nyquist sampling. The synthesized beam is located at the bottom right corner. Polarization intensity and B-segments are only presented where the polarization intensity shows a $>$3$\sigma$ detection. }
  \label{fig:continuum}
\end{figure}
The upper panel of Figure~\ref{fig:continuum} shows the 870 $\mu$m dust continuum emission image. Overplotted are associated infrared and radio sources, as well as OH and H$_{2}$O masers.
This continuum image reveals asymmetric dense gas structures. 
They form a well resolved $\sim0.07$~pc ($\sim$14,000 AU) scale arc-like structure along the east-west direction (measured from the $8\sigma$ contour from the 870~$\mu$m dust continuum emission), with three distinguishable peaks surrounding a lower density cavity, labeled as N (North), SE (South-East) and HC (Hot Core; see Section~\ref{sec:molecules}). 
Positions, deconvolved sizes, peak intensities, flux densities and mass estimates of the three main peaks are listed in Table~\ref{tab:dustpar}. 
Most of the known infrared and radio sources are located either around the 870 $\mu$m peak (hot core; e.g., IRS V3 and radio source R E-3), or are inside the cavity.

Based on the VLA observations of the centimeter free-free continuum emission presented in \citet{Rengarajan96}, we constrain the 870 $\mu$m free-free emission to be < 4.7 mJy towards R-E1, R-E2 and R-E3, which is negligible as compared to the detected emission level in Figure~\ref{fig:continuum} (see table~\ref{tab:dustpar}). Assuming that the 870 $\mu$m dust emission is optically thin, we estimated the gas+dust masses according to 
\begin{equation}
M=R\frac{d^2S_\nu}{B_\nu(T_d)\kappa_\nu},
\end{equation} 
\begin{table*}
\caption{Parameters of the sources detected with the SMA at 870~$\mu$m dust continuum emission.}
\begin{center}
{\small
\begin{tabular}{lcccccccc}
\hline
\hline
&\multicolumn{2}{c}{Position$^\mathrm{a}$}
&Deconv.ang.size$^\mathrm{a}$
&P.A.$^\mathrm{a}$
&$I_\mathrm{\nu}^\mathrm{peak}$~$^\mathrm{a}$
&$S_\mathrm{\nu}$~$^\mathrm{a,b}$
&Mass$^\mathrm{c}$
\\
\cline{2-3}
Source
&$\alpha (\rm J2000)$
&$\delta (\rm J2000)$
&$(''\times'')$
&($^\circ$)
&(Jy Beam$^{-1}$)
&(Jy)
&(M$_\odot$)\\
\noalign{\smallskip}
\hline\noalign{\smallskip}
HC	   	&17:19:57.364 &$-35$:57:53.47	&$2.9\pm0.6\times 1.66\pm0.6$ 	&$-89\pm41$		&$1.27\pm0.16$ 		&$3.6\pm0.7$ 		&4-9 \\ 		
N core 	&17:19:57.408 &$-35$:57:48.55	&$6.2\pm2.3\times 2.62\pm1.2$ 	&$85\pm18$		&$0.50\pm0.02$ 		&$3.3\pm0.7$ 		&16 \\
SE core 	&17:19:57.792 &$-35$:57:53.50	&$2.4\pm0.4\times 1.78\pm0.3$ 	&$45\pm20$		&$0.80\pm0.04$ 		&$1.9\pm0.4$ 	 	&9\\
\hline
\label{tab:dustpar}
\end{tabular}
\begin{list}{}{}
\item[$^\mathrm{a}$] Position, deconvolved size, position angle (P.A.), peak intensity, flux density and the respective uncertainties derived from fitting a 2D Gaussian to each source using the {\sc imfit} task from {\sc Miriad}. Peak intensities and flux densities are corrected for the primary beam response.
\item[$^\mathrm{b}$] Error in flux density is calculated as $\sqrt{(\sigma\,\theta_\mathrm{source}/\theta_\mathrm{beam})^2+(\sigma_\mathrm{flux-scale})^2}$ \citep{Beltran01,Palau13}, where $\sigma$ is the rms of the map, $\theta_\mathrm{source}$ and $\theta_\mathrm{beam}$ are the sizes of the source and the beam, respectively, and $\sigma_\mathrm{flux-scale}$ is the error in the flux scale, which takes into account the uncertainty on the calibration applied to the flux density of the source ($S_\nu\times\%_\mathrm{uncertainty}$) which we assumed to be 20\%. 
\item[$^\mathrm{c}$] Masses derived assuming a dust mass opacity coefficient of 2.03 cm$^{-2}$g$^{-1}$ \citep[for thin ice mantles after $10^5$ years of coagulation at a gas density of $10^6$ cm$^{-3}$,][]{Ossenkopf94}, a dust temperature of 50 and 100~K for the hot core (HC) and 30 K for the north (N) and southeast (SE) cores. The uncertainty in the masses due to the opacity law and temperature is estimated to be a factor of 4.
\end{list}
}
\end{center}
\end{table*} 
where $d$ is the distance, $S_\nu$ is the flux density, $B_\nu(T_d)$ is the Planck function at the isothermal dust temperature $T_d$, $\kappa_\nu$ is the dust opacity and the factor R is the gas-to-dust ratio which we assumed to be 100 \citep{Hildebrand83,Lis98}. We adopted $\kappa_\nu=2.03$~cm$^2$g$^{-1}$ for thin ice mantles after $10^5$ years of coagulation at a gas number density of $10^6$~cm$^{-3}$ and for the frequency of our observations \citep{Ossenkopf94} and we assumed $T_d=50-100$~K \citep[hot core temperatures $\gtrsim100$ K; e.g.,][]{Palau11} for the hot core and 30~K for the north (N) and southeast (SE) cores \citep[the average temperature in protostellar cores, e.g.,][see Section~\ref{outflow}]{Faundez04,Sanchez-Monge13}. 
The resulting masses are given in Table~\ref{tab:dustpar}.
The mass derived from the total 870 $\mu$m flux density detected by the SMA is $\sim50$~M$_{\odot}$ for a $T_d=30$~K. This value is <10\% of the mass estimated by \citet{Munoz07} based on 1.2~mm dust continuum emission obtained with the SEST SIMBA bolometer array. There are two reasons for this 
difference. We might be missing some flux due to the interferometric filtering of extended emission, and \citet{Munoz07} use an integration area for the flux calculation approximately ten times larger than our work (using an effective radius of 0.86 pc).

\subsubsection{Polarization data}	\label{sec:polarization}
\begin{figure}
 \centering
\includegraphics[scale=0.3,keepaspectratio=true,angle=-90]{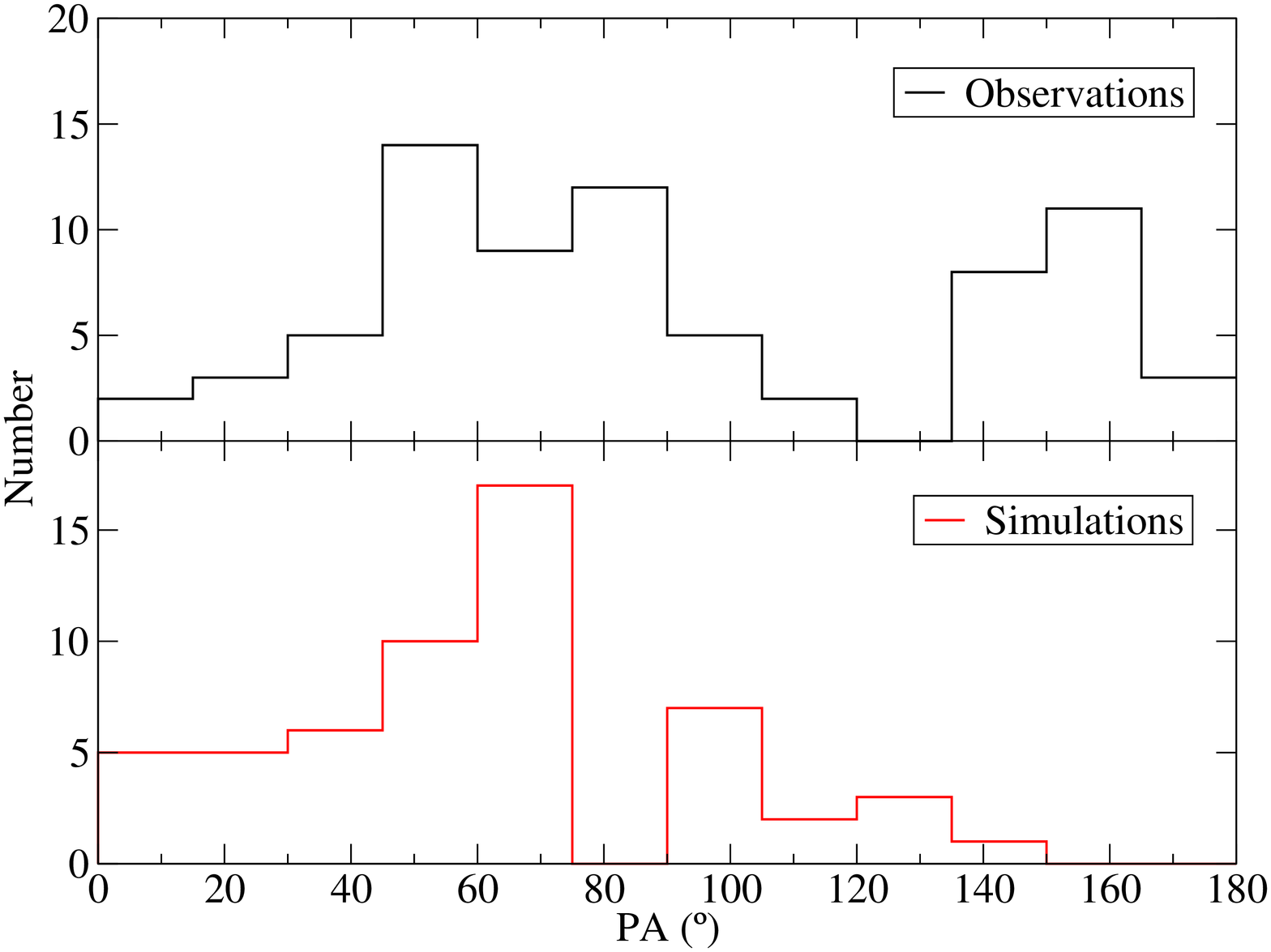}
  \caption{{\it Upper panel}: Distribution of position angles of the magnetic field segments shown in Fig.~\ref{fig:continuum} for a polarized emission cutoff of 3$\sigma$. {\it Lower panel}: Distribution of position angles of the magnetic field segments produced by the simulations (see lower panel of Fig.~\ref{fig:modelpanels}).}
  \label{distPA}
\end{figure}

It is generally accepted that non-spherical interstellar dust grains are locally aligned with the B-field, with their longest axes being perpendicular to the B-field directions \citep[see][for a review]{Hildebrand00}.
Therefore, the projected orientation of the B-field can be inferred by rotating the dust linear polarization (i.e., electric field) by 90 degrees (hereafter B-segment for the 90$^{\circ}$ rotated linear polarization).

The lower panel of Figure~\ref{fig:continuum} shows the polarization map from the 870 $\mu$m continuum observations. The projected magnetic field shows an east-west orientation following the dust morphology, from the south-east (SE) continuum peak towards the hot core region. There are also several detections at the northern part of the dust continuum emission showing a near north-south orientation. This northern component of the magnetic field seems to connect to the hot core position through several detections presenting a northeast-southwest orientation.  

The histogram of the B-segment position angles (see upper panel of Figure~\ref{distPA}) shows a disperse distribution due to the changes in the morphology of the magnetic field. We identified two major orientations, at $\sim90^\circ$ (east-west component) and $\sim50^\circ$ (northeast-southwest component) which seem to be converging at the hot core region. The northern component at $\sim160^\circ$ (north-south) is also apparent as a separate peak. 

\subsubsection{Molecular lines} \label{sec:molecules}
\begin{table*}
\caption{Molecular lines}
\begin{center}
\begin{tabular}{c c c c}
\hline
\hline
Molecular line		&Transition	 		&Frequency (GHz)	&E$_{\text{U}}$ (K) 	\\
\hline
Hot core tracers\\
\hline
SO$_2$			&21(2,20)$-$21(1,21)		&332.09143		&220			\\
SO$_2$			&16(4,12)$-$16(3,13)		&346.52388		&164			\\
SO$_2$			&19(1,19)$-$18(0,18)		&346.65217		&168			\\
$^{34}$SO$_2$	&16(4,12)$-$16(3,13)		&332.83622 		&163			\\
$^{34}$SO$_2$	&19(1,19)$-$18(0,18)		&344.58104		&168			\\
$^{34}$SO$_2$	&11(4,8)$-$11(3,9)			&344.99816		&99				\\
$^{34}$SO$_2$	&8(4,4)$-$8(3,5)			&345.16866		&71				\\
$^{34}$SO		& 7(8)$-$6(7)				&333.90098 		&80				\\
CH$_3$OH		&18(2,16)$-$17(3,14)		&344.10913 		&419			\\
CH$_3$OH		&19(1,19)$-$18(2,16)++		&344.44390		&451			\\
CH$_3$OCHO		&31(0,31)$-$30(1,30)		&333.44902 		&260			\\
CH$_3$OCH$_3$	&19(1,19)$-$18(0,18)AA		&344.35806		&167			\\
CH$_3$OCH$_3$	&11(3,9)$-$10(2,8)EE		&344.51538		&73				\\
HC$_3$N			&J=38$-$37					&345.60901		&323			\\

\hline
Dense core tracers\\
\hline
SO				&8(8)$-$7(7)				&344.31061		&87				\\
SO$_2$ 			&4(3,1)$-$3(2,2)			&332.50524		&31 			\\
SO$_2$			&13(2,12)$-$12(1,11)		&345.33854		&93				\\
CH$_3$OH		&7(1,7)$-$6(1,6) 			&335.58200 		&79				\\
HC$^{15}$N 		&(4$-$3)					&344.20011		&41				\\
NS				&J=15/2$-$13/2				&346.22116		&71				\\
H$^{13}$CO$^+$	& (4$-$3)					&346.99835		&41				\\
\hline
Outflow tracers\\
\hline
CO 				&(3$-$2)					&345.7959		&33				\\
SiO				&(8$-$7)					&347.33058		&75				\\
\hline		
\end{tabular}
\end{center}
\label{transitions}
\end{table*}
\begin{figure}
 \centering
  \includegraphics[scale=0.6,keepaspectratio=true]{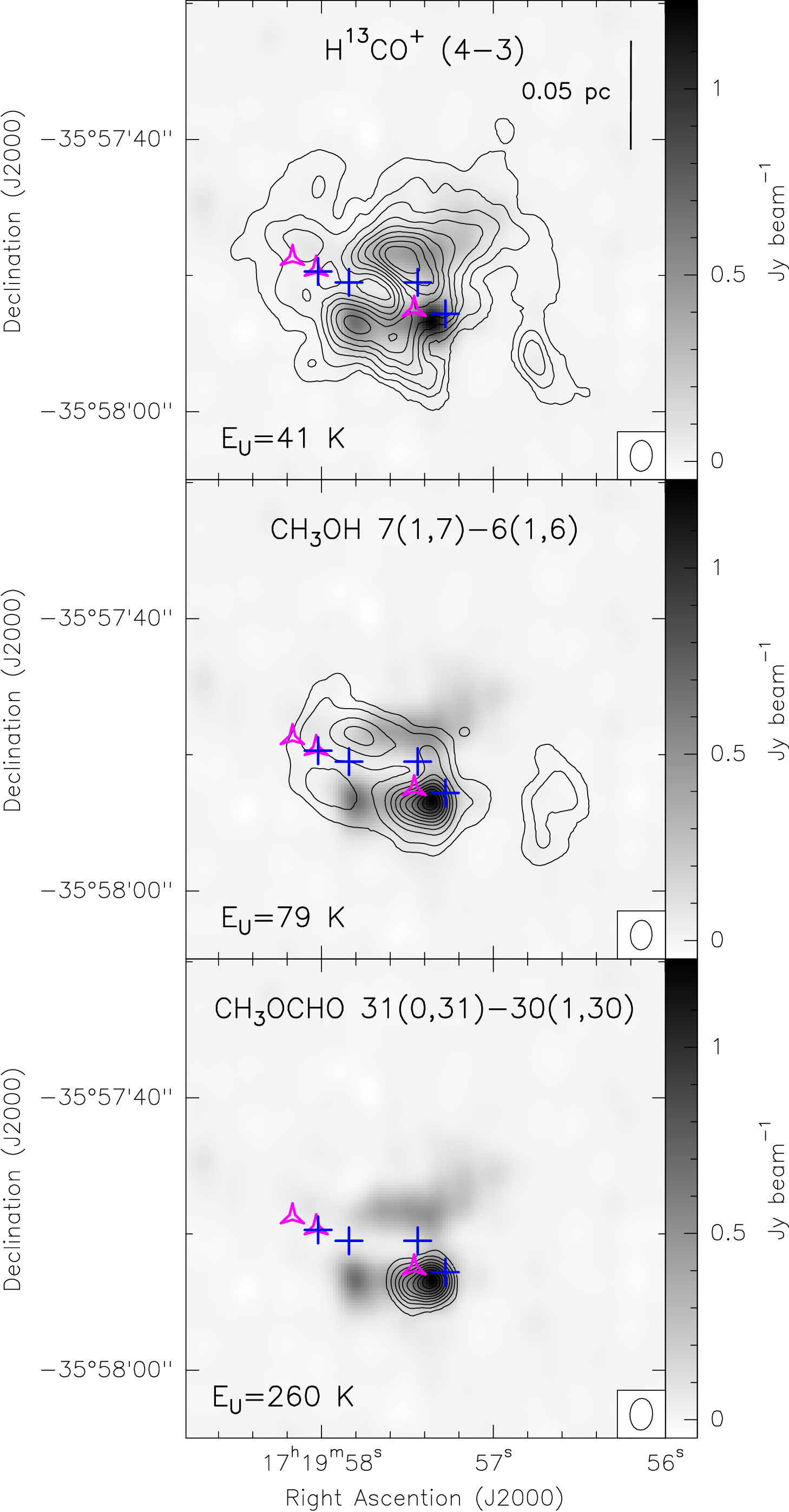}
  \caption{Grey scale image of the 870 $\mu$m dust continuum emission overlaid with the contour maps of the velocity integrated emission (i.e., moment 0 maps) of selected molecular line tracers. {\it Upper panel}: Contours: H$^{13}$CO$^+$ (4$-$3). Contour levels are 10, 20, 30, ..., 90 percent of the peak value, 11.2 Jy beam$^{-1}$ km s$^{-1}$. The synthesized beam located at the bottom right is $2\farcs28\times1\farcs57$, P.A.$=-2.72^\circ$. {\it Middle panel}: Contours: CH$_3$OH 7(1,7)$-$6(1,6). Contour levels are  5, 10, 20, 30, ..., 90 percent of the peak value, 37.1 Jy beam$^{-1}$ km s$^{-1}$. The synthesized beam located at the bottom right is $2\farcs19\times1\farcs54$, P.A.$=-3.45^\circ$. {\it Lower panel}: Contours: CH$_3$OCHO 31(0,31)$-$30(1,30). Contour levels are  10, 20, 30, ..., 90 percent of the peak value, 11.1 Jy beam$^{-1}$ km s$^{-1}$. The synthesized beam located at the bottom right is $2\farcs35\times1\farcs62$, P.A.$=-3.17^\circ$. Blue crosses and pink triangles are labeled as in Fig~\ref{fig:continuum}.}
 \label{fig:mom0}
\end{figure}
\begin{figure}
 \centering
 \includegraphics[scale=0.4,keepaspectratio=true]{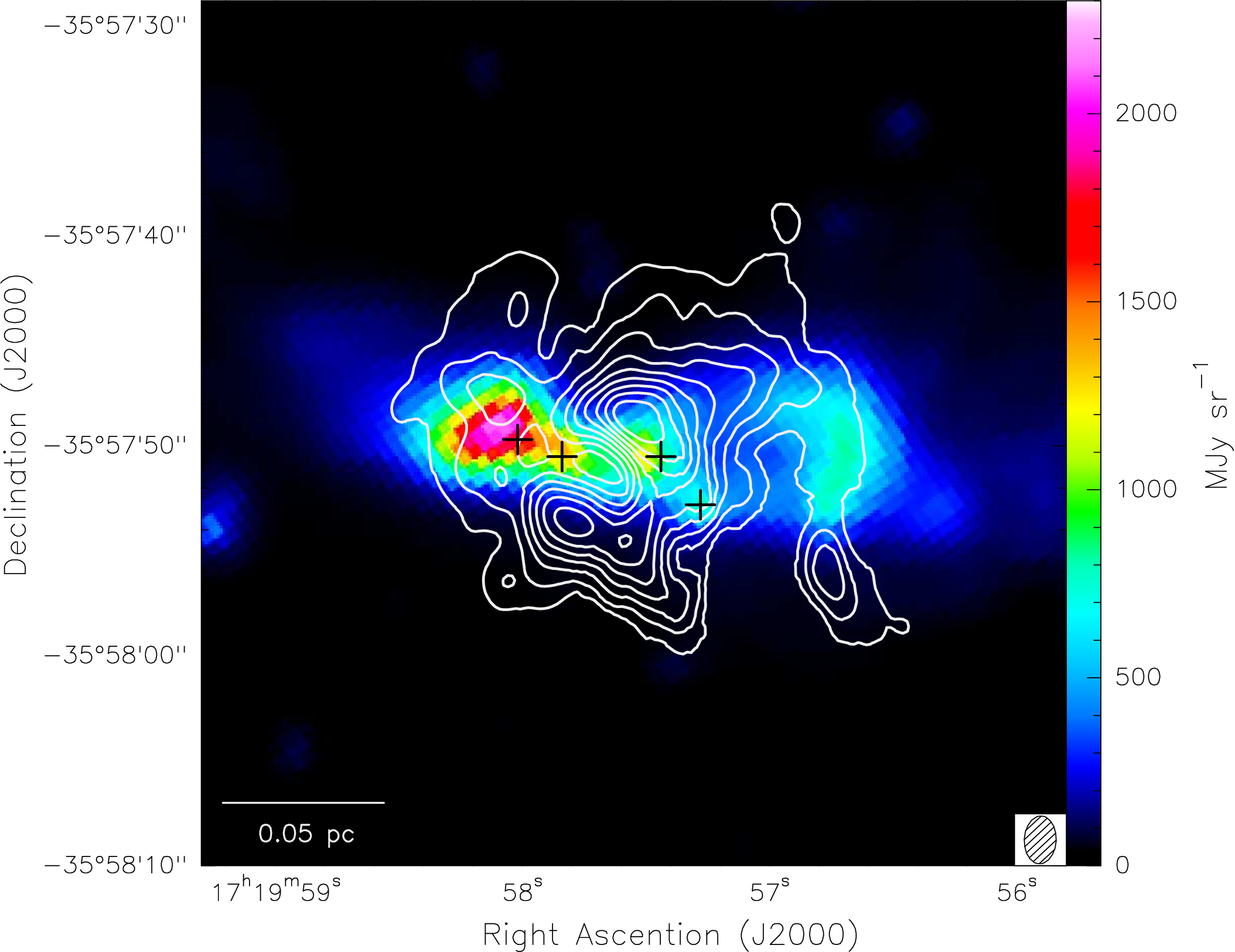}
  \caption{The Spitzer 4.5 $\mu$m image (color scale), and the velocity integrated emission (moment 0) map of H$^{13}$CO$^+$ (4$-$3) (contour). The synthesized beam located at the bottom right corner is $2\farcs28\times1\farcs57$, P.A.$=-2.72^\circ$. Black crosses show the positions of infrared sources \citep{Kraemer99}.}
      \label{spitzer}
\end{figure}
We have detected 24 molecular lines in the 8 GHz band.
Table~\ref{transitions} lists the frequency and the energy of the upper level (E$_\mathrm{U}$) for each transition. We found that the emission from several molecular transitions with E$_\mathrm{U}$ between $\sim100-400$ K spatially coincides with the strongest dust continuum peak (see Fig.~\ref{fig:mom0} and Fig.~\ref{fig:mom1}), including transitions from SO$_2$, $^{34}$SO$_2$, HC$_3$N, CH$_3$OH, CH$_3$OCH$_3$ and CH$_3$OCHO. Thus, these molecular transitions likely trace a hot molecular core.

In Fig.~\ref{fig:mom0} we present the moment 0 (integrated intensity) maps of H$^{13}$CO$^{+}$ (4--3), CH$_3$OH 7(1,7)$-$6(1,6), and CH$_{3}$OCHO 31(0,31)$-$30(1,30), which have upper level energies of 41 K, 79 K, and 260 K, respectively. The very different excitation conditions for these molecular line transitions trace different regions in the molecular gas, which show a variety of morphologies.  
H$^{13}$CO$^+$ (4$-$3) (upper panel) is a good tracer of the dense and relatively cold gas and traces the extended dense core of NGC 6334 V. 
It shows two main peaks forming a ring-like structure with lack of emission at the center of the ring, coinciding with the dust continuum cavity. 
CH$_3$OH 7(1,7)$-$6(1,6) (middle panel) traces warmer gas, and it is generally a good hot core tracer. 
It presents strong emission at the hot core position and also traces a more compact region of the dense core, forming a ring-like structure, as the H$^{13}$CO$^+$ molecular line. 

Many of the infrared and radio sources found in the region are located where we find a deficit of molecular emission in both H$^{13}$CO$^+$ and CH$_3$OH (center of the ring).
In addition, the cavity is filled with 4.5 $\mu$m emission (see Fig.~\ref{spitzer}). The origin of this extended emission is unclear. Possible emission mechanisms could be scattered continuum emission in outflow cavities \citep[e.g.,][]{Qiu08,Takami12}, H$_2$ emission and CO emission from shocks \citep[e.g.,][]{Noriega-Crespo04,Smith06,Davis07,Takami12}. In any case, the 4.5 $\mu$m emission is likely to be a signpost of a cavity generated by a still active outflow.

Finally, in the lower panel of Fig.~\ref{fig:mom0}, we present a high-excitation transition (E$_U$=260 K) of CH$_3$OCHO which shows more compact emission elongated along an east-west direction, centered only at the hot core region.

\subsection{Kinematics}
A Gaussian fit to the averaged spectrum of the entire emission of the dense core tracer H$^{13}$CO$^+$ (4$-$3) suggests that the systemic velocity of NGC 6334 V is $-5.7$~km~s$^{-1}$. Therefore, we use this value for the following kinematic analysis.
\subsubsection{Velocity gradients and mass estimates} \label{velocity}
\begin{figure}
 \centering
  \includegraphics[scale=0.6,keepaspectratio=true]{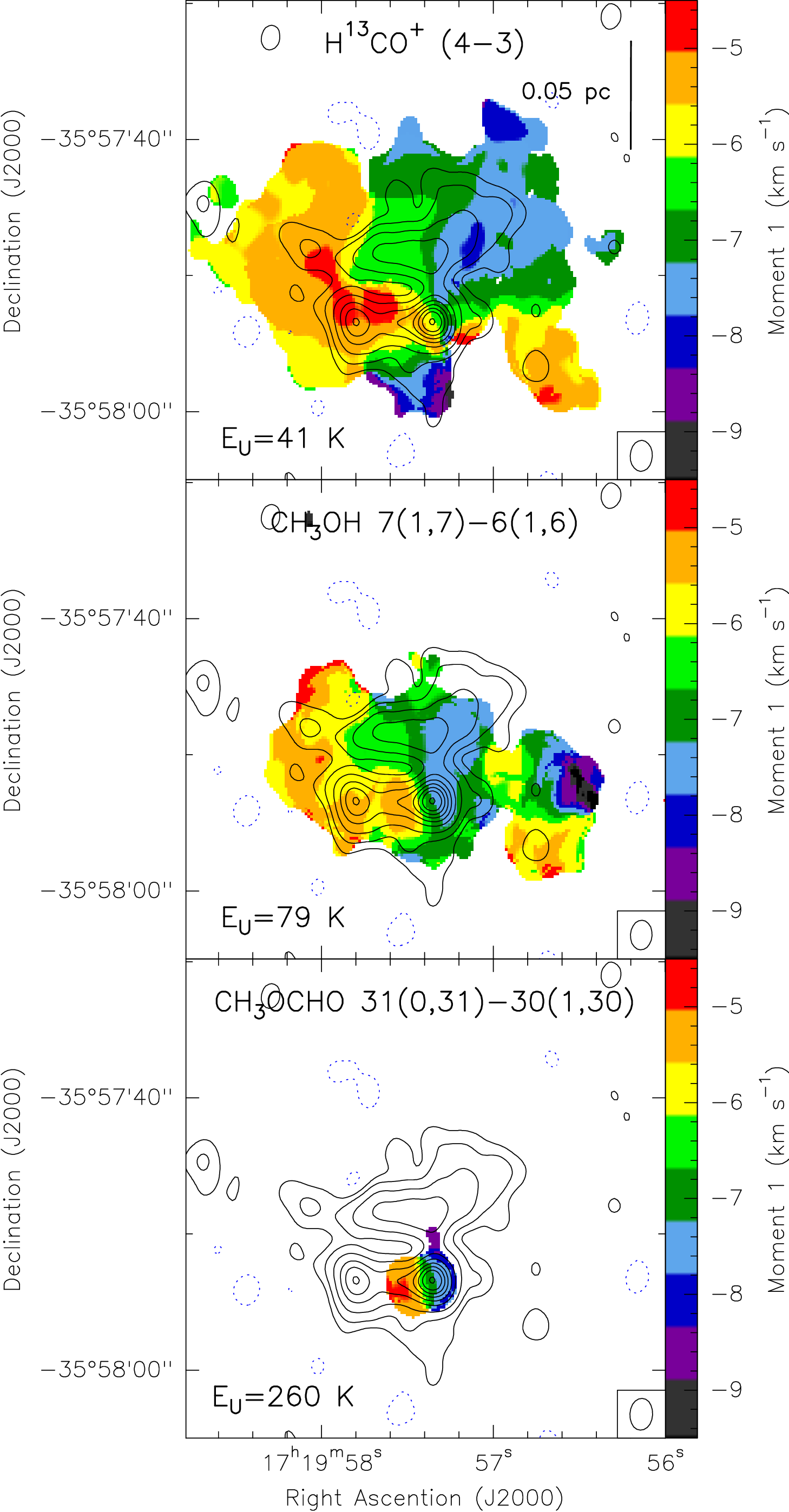}
  \caption{Contour map of the 870 $\mu$m dust continuum emission overlaid with the intensity-weighted average velocity (i.e., moment 1) images (color scale) of the following molecular lines: H$^{13}$CO$^+$ (4$-$3) (upper panel), CH$_3$OH 7(1,7)$-$6(1,6) (middle panel) and of CH$_3$OCHO 31(0,31)$-$30(1,30) (bottom panel). Synthesized beams of the molecular line images are shown in the bottom right corners of the panels.}
 \label{fig:mom1}
\end{figure}
A comparison between the different gas tracers can provide clues about the kinematics from larger (extended dense core) to smaller (hot core) scales. Figure~\ref{fig:mom1} shows the intensity-weighted averaged velocity maps (i.e., moment 1 maps) of the same molecular lines shown in Fig.~\ref{fig:mom0}. 
The extended dense core tracers H$^{13}$CO$^+$ and CH$_3$OH (upper and middle panels, respectively) show a velocity field dominated by a clear
east-west gradient from $-8$ to $-5$ km s$^{-1}$. However, the overall velocity field has a quadrupolar pattern centered at the hot core region. This excludes a simple interpretation for the velocity field, such as a rotating core.

To know if the observed gas belongs to a gravitationally bound system, we estimated the needed minimum mass associated to the observed velocity gradient. The maximum velocity for an object accelerated in a gravitationally bound system is given by $v_{\mathrm{max}}=(2GM/R)^{1/2}$ where $G$ is the gravitational constant, $R$ is the distance to the gravitational center and $M$ is its mass. Adopting the gravitational center as the center of the velocity gradient located at the hot core position and using a distance of $6''$ and a velocity of 2~km~s$^{-1}$ from the observed velocity gradient, we derive a mass of $\sim18$ M$_{\odot}$.  Note that the radial velocity is a lower limit of the total velocity. This mass is lower than the mass estimated from the dust continuum emission, which is a lower limit considering that we are filtering extended emission with the SMA (see Section~\ref{continuum}), and thus, we consider the system to be gravitationally bound. 

The hot core tracer CH$_3$OCHO also presents a clear velocity gradient following the orientation of the larger-scale velocity field. The intermediate velocity $-6.5$ km s$^{-1}$ is shifted from the dust peak position $0.5''$ or 650 AU towards the east (several transitions of CH$_3$OH, CH$_3$OCH$_3$, $^{34}$SO$_2$ and SO$_2$ also present the same velocity field). The velocity gradient is also compatible with a gravitationally bound system, with an enclosed mass of $\sim6$ M$_{\odot}$ (using $R=2''$ and $v=2$~km~s$^{-1}$). The lower limit of the hot core mass obtained from the dust continuum emission is $\sim$4--9 M$_{\odot}$. 

\subsubsection{High-velocity gas} \label{outflow}
\begin{figure*}
 \centering
 \includegraphics[scale=0.6,keepaspectratio=true]{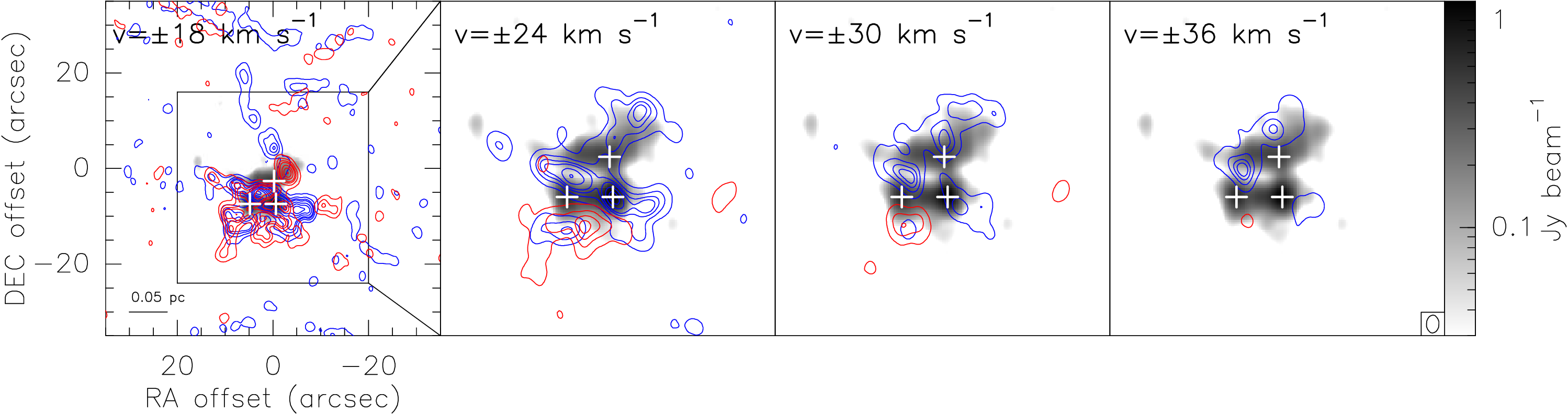}
  \caption{Velocity channel maps of CO (3--2) (red and blue contours), overplotted on the 870 $\mu$m dust continuum image (gray scale). Individual panels present emission at various velocities relative to the systemic velocity $v_{lsr}$=$-5.7$ km s$^{-1}$. Contour levels are $-3$, 3, 9, 15, 20, 30, 40, ..., 70 times the rms noise level 90~mJy~beam$^{-1}$. The synthesized beam located at the bottom right corner of the last panel is $2\farcs13\times1\farcs50$, P.A.$=-2.16^\circ$. The white crosses indicate the three dust continuum peaks (N, SE and HC). A scale bar is located at the left bottom of the first panel. The black square in the first panel indicates the field of view of the rest of the panels.}
  \label{COpanels}	
\end{figure*}
Figure~\ref{COpanels} shows the red- and blue-shifted CO (3$-$2) emission at velocities of $\pm18$, $\pm24$, $\pm30$ and $\pm36$~km~s$^{-1}$ from the systemic velocity, $-5.7$~km~s$^{-1}$. At these high velocities, the CO (3$-$2) is tracing molecular outflows. The first panel ($\pm18$~km~s$^{-1}$) shows a very complicated structure with red- and blue-shifted emission spread all over the dust continuum. In the second panel ($\pm24$~km~s$^{-1}$), we can distinguish collimated blue-shifted emission along the dust continuum cavity. This emission suggests that the high-velocity gas may have produced the observed cavity. Furthermore, in this panel, we observe blue- and red-shifted emission extended towards the north and south of the hot core position, respectively, in agreement with the larger-scale previously reported CO outflow \citep{Kraemer95,Zhang14}. This emission is oriented almost perpendicular to the velocity gradient seen at the hot core region (see bottom panel of Fig.~\ref{fig:mom1}). Note that outflow powering source candidates, KDJ3 and KDJ4 \citep{Kraemer99,Hashimoto07,Simpson09}, are located close to the hot core position (see Fig.~\ref{fig:continuum}). KDJ3 does not have dust continuum emission associated as it is located near the dust cavity. On the other hand, KDJ4 seems to be a younger object than KDJ3 and a better outflow powering source candidate as it coincides well with the hot core dust continuum peak. 

The highest outflow velocities, at $\pm30$ and $\pm36$~km~s$^{-1}$ in the last two panels of Fig.~\ref{COpanels}, show red-shifted emission near the southeast continuum peak (SE) and blue-shifted emission towards the hot core region, the northern continuum peak and the dust cavity. Especially in the third panel (at $\pm30$~km~s$^{-1}$), it seems that the southeast (SE) and north (N) cores are associated with high-velocity gas. Thus, even though the CO high-velocity emission presents a very complicated structure, it suggests the presence of at least two or three independent outflows. Multiple molecular outflows associated with a massive dense core have been previously reported in several high-mass star-forming regions \citep[e.g.][]{Naranjo-Romero12,Girart13,Fernandez-Lopez13}.

\section{Analysis}
\label{sec:analysis}
\subsection{Converging flows}

\begin{figure}
\centering
\begin{minipage}[c]{0.75\linewidth}
\includegraphics[width=\textwidth]{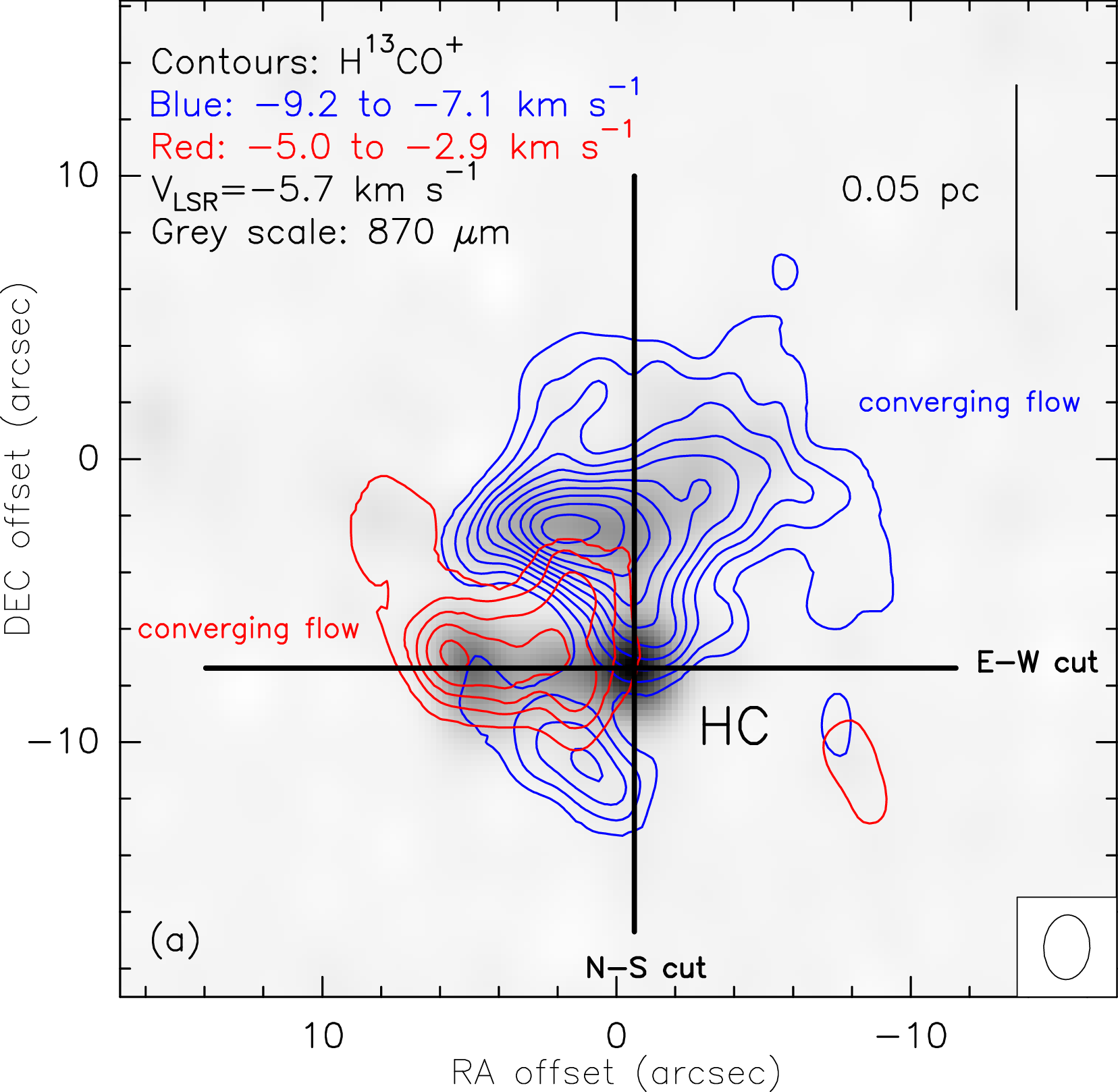}
\end{minipage}\hfill
\\
\begin{minipage}[l]{0.99\linewidth}
\includegraphics[width=\textwidth]{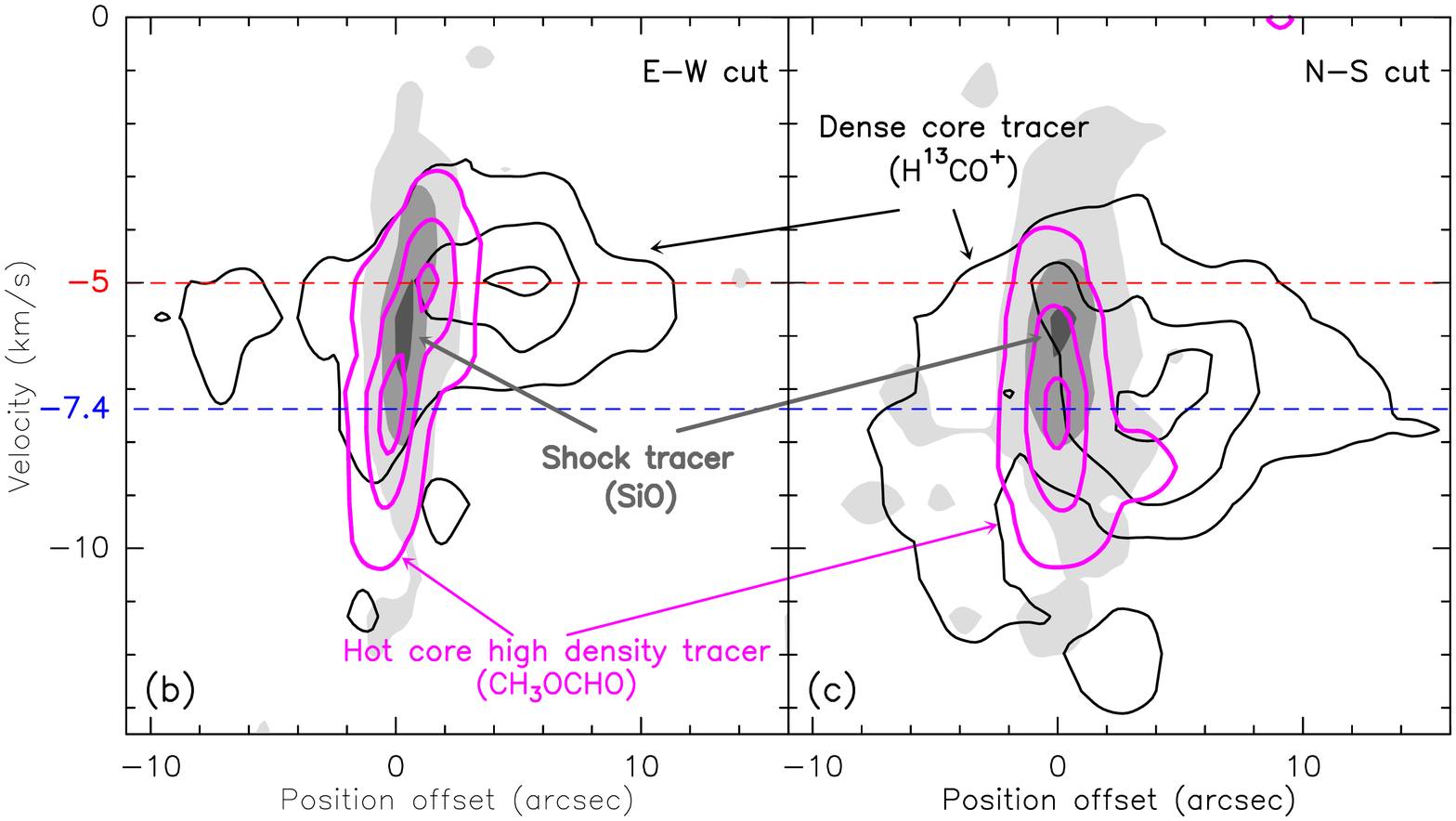}
\end{minipage}\hfill
\caption{Panel (a): Blue and red contours: averaged emission from blue and redshifted channels of H$^{13}$CO$^+$\,(4--3) displaying the two flow components. Contours are 4, 8, 12, 16, 20 times the rms noise level of the map, 60 mJy beam$^{-1}$. The synthesized beam is located at the bottom right corner. Black solid lines show position-velocity cuts that are shown in panels (b) and (c). We can see that H$^{13}$CO$^+$ does not trace the hot core as seen also in other studies (e.g., Girart et al. 2013). Grey scale: 870 $\mu$m dust continuum emission. (b): E-W position velocity cuts along the H$^{13}$CO$^+$\,(4--3) (blue), CH$_3$OCHO 31(0,31)$-$30(1,30) (pink) and SiO\,(8--7) (grey scale) line emission passing through the hot core. Contours are 10, 50 and 90 percent of the peak values 4.0 Jy beam$^{-1}$ (H$^{13}$CO$^+$), 2.9 Jy beam$^{-1}$ (CH$_3$OCHO) and 1.5  Jy beam$^{-1}$ (SiO). (c): same as panel (b) but along N-S position velocity cuts. Dashed horizontal lines indicate the two velocity components at $-5$ and $-7.4$~km~s$^{-1}$. 
\label{fig:pvcuts}}
\end{figure}

As we have seen in Section~\ref{velocity}, the molecular dense core traced by H$^{13}$CO$^+$ and CH$_3$OH presents a velocity gradient with a certain symmetry centered at the hot core region (see upper and middle panels in Fig.~\ref{fig:mom1}). Figure~\ref{fig:pvcuts}a shows the blue- and red-shifted integrated emission of H$^{13}$CO$^+$ (4--3) showing two filamentary structures at the eastern and northern parts of the hot core. Both velocity structures seem to converge to the hot core. At the hot core region, the velocity gradient follows the orientation of the larger-scale velocity field (see Fig.~\ref{fig:mom1}). As the center of the velocity gradient is shifted from the dust peak, the observed velocity gradient is not likely due only to rotational motions but additional motions may be also affecting it (e.g., infalling motions).

In Figure~\ref{fig:pvcuts}b and c we show two position-velocity diagrams, along the east-west and north-south direction centered at the hot core region, to compare the kinematics from the extended dense core (traced by H$^{13}$CO$^+$) and the hot core region (traced by CH$_3$OCHO). Both dense core and hot core tracers show the same two distinctive velocity components, one at $-5.0$ km s$^{-1}$ and another one at $-7.4$ km s$^{-1}$. In addition, we did the same position-velocity cuts towards the shock tracer SiO\,(8--7). The SiO emission is a signpost of strong shocks. It has, however, also been detected at velocities close to the systemic velocity without a clear association to protostellar outflow activity \citep{Jimenez-Serra10,Nguyen-Luong13}. In these cases, SiO can be produced in low-velocity shocks \citep[e.g.][]{Duarte-Cabral14,Girart16}.
As shown in grey scale in Fig.~\ref{fig:pvcuts}b and c, SiO presents a velocity component just at an intermediate velocity of $\sim-6.0$ km s$^{-1}$ and at the dust continuum peak position, suggesting that there could be interaction between the two filamentary structures precisely at the hot core region. Such a converging flow toward a hot core was also reported in W33A by \citet{Galvan-Madrid10}.

\begin{figure}
 \centering
 \includegraphics[scale=0.53,keepaspectratio=true]{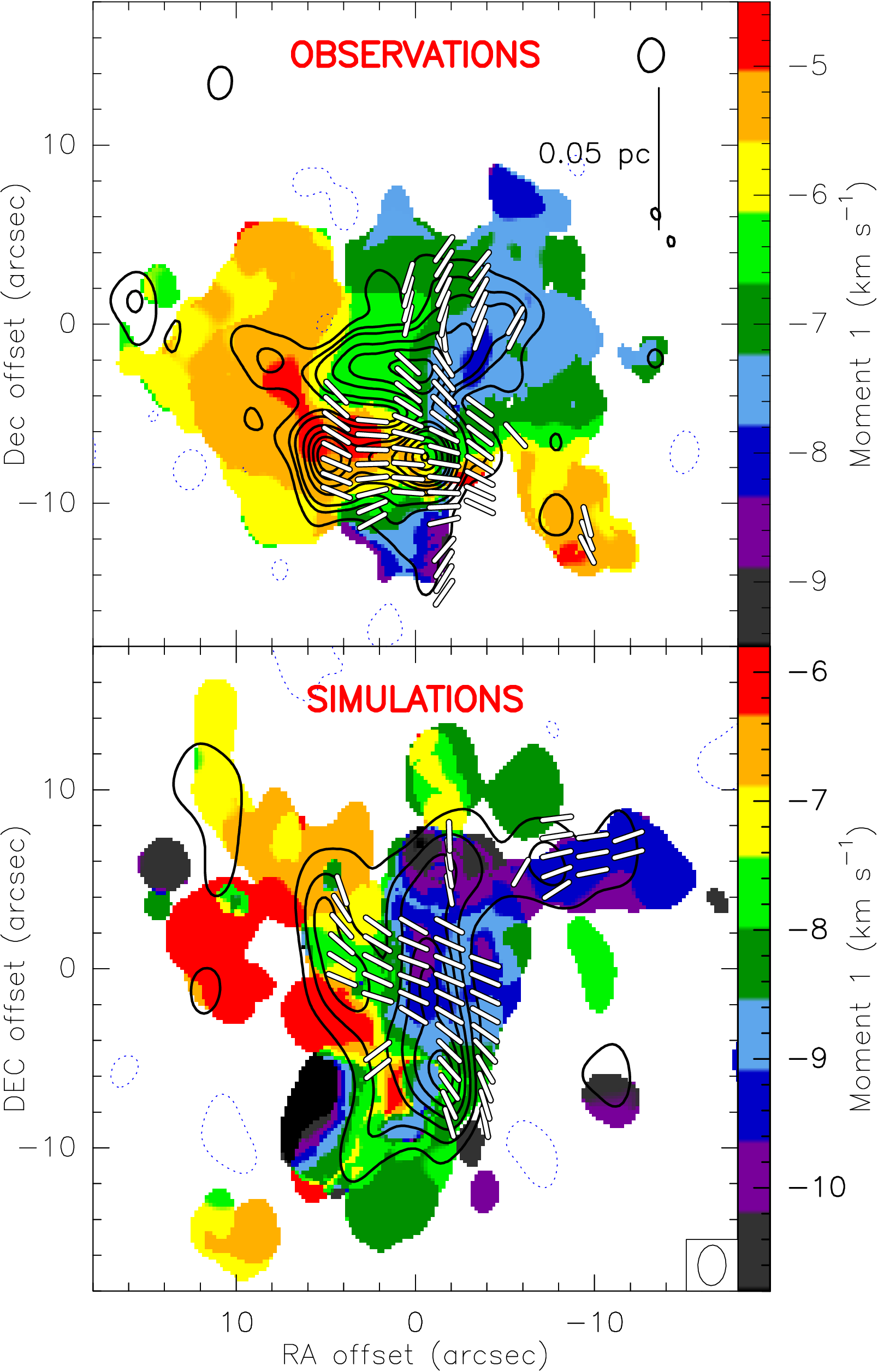}
  \caption{A comparison of the observed dense gas distribution, the velocity field, and the B-segment orientations, with the simulated ones. {\it Upper panel}: SMA observations. White segments: Magnetic field orientations projected on the plane of the sky (only compact and subcompact configurations). Color scale: Moment 1 map of H$^{13}$CO$^+$ (4$-$3). Black contours: 870 $\mu$m dust continuum emission. {\it Lower panel}: Simulations convolved with SMA response. Color scale: H$^{13}$CO$^+$ (4$-$3) moment 1 map (velocity field) from ARTIST. The synthesized beam is located at the bottom right corner. White segments: Magnetic field orientations obtained with the DustPol module of ARTIST. Black contours: Simulated 870 $\mu$m dust continuum emission.}
      \label{fig:modelpanels}
\end{figure}

Furthermore, the magnetic field distribution described in Section~\ref{sec:polarization} also shows a ``bimodal converging'' pattern towards the hot core. In the upper panel of Figure~\ref{fig:modelpanels} we overlaid the moment 1 map of H$^{13}$CO$^+$ with the magnetic field distribution. We used the polarization detections from only compact and subcompact configurations as the polarized emission is stronger than with all three configurations combined. The morphology of the magnetic field is very well correlated with the two velocity components shown by the H$^{13}$CO$^+$ velocity field, suggesting that the magnetic field could be being dragged by the gas dynamics. Previous observations of the B-type binary forming region G192.16-3.84 \citep{LiuHB13} have shown that the magnetic field is also dragged by the gravitationally dominant infall and rotational motions, where the magnetic field orientation is approximately parallel to the velocity gradient of the gas.

The combination of these results shows how the gas seems to be infalling from the larger-scale dense core of NGC 6334 V towards the higher-density hot core region through two distinctive converging flows dragging along the magnetic field whose strength seems to have been overcome by gravity.


\subsection{Synthetic observations}
Our observational analysis so far points out that the gas is converging towards the hot 
core through two main infalling flow components, suggesting that gravity could play 
a major role in the dynamics of intermediate/massive star-forming regions. 
However, our interpretation can be biased due to several observational effects. 
First, certain information has been lost when projecting a 3-dimensional density distribution onto a 2-dimensional plane. 
In addition, the vector fields (e.g., B-field, mean velocity, etc) were smeared due to the integration along the line-of-sight.
Moreover, the interferometric observations of molecular lines are biased by the excitation conditions and the incomplete sampling
in the Fourier domain.
To facilitate the test of whether or not the observational results are indeed consistent with our interpretation,
we have carried out 3-D numerical magnetohydrodynamic (MHD) simulations for massive star-forming regions which are dominated by gravity.
We produced synthetic observational results based on these simulations, to make a consistent comparison with the observational results of SMA.

\subsubsection{Numerical simulation}
We carry out a 3-D numerical simulation of a massive, compact molecular
cloud, similar to that presented in \citet{Ballesteros-Paredes15}, although with a
different numerical scheme: rather than the Lagrangian smoothed particle
hydrodynamical code used before, we use the Eulerian adaptive mesh
refinement (AMR) FLASH (version 2.5) code \citep{Fryxell00} in a
magneto-hydrodynamical regime.  As in \citet{Ballesteros-Paredes15}, we include 
self-gravity, and the formation of sink particles, in an initially
turbulent velocity field which is let to evolve freely, i.e., no
turbulent forcing is imposed.  The simulation is isothermal in order
to isolate the thermal effects on the cloud evolution. We also did not
include radiative feedback or stellar heating from the sink
particles in order to understand the very early stages of the
molecular cloud collapse.

The initial conditions are as follows. We consider a numerical box of 1~pc per side containing 1000~M$_\odot$
of molecular gas. The mass in the box is distributed homogeneously,
with a number density of $n_0=1.7\times10^4 \, \mathrm{cm}^{-3}$. The free-fall time for this
configuration is 256,000~yr.  Following the prescription by, e.g.,  
\citet[][]{Stone98}, we included a pure rotational velocity
spectrum with random phases and amplitudes that peak at wave numbers
$k= 4\pi/L_0$, where $L_0$ is the linear size of the box. The
resulting initial velocity field was, thus, an incompressible supersonic
turbulent fluid, with an rms Mach number of $\mathcal{M}_{\rm rms}=8$.
No forcing at later times is imposed.

The magnetic field is initially uniform along the $x$-direction with a 
strength of $50 \, \mu \mathrm{G}$, which is
consistent with magnetic field intensities at densities of the order
of $n_0$ \cite[see e.g.,][]{Crutcher12}.  Our box is, thus,
magnetically supercritical, with a mass-to-flux ratio, $\mu$, 6.7
times the critical value, $\mu_\mathrm{crit}=(4 \pi^2 G)^{-1/2}$
\cite[see e.g.,][]{Nakano78}, and prone to gravitational collapse as soon
as the initial turbulent field is dissipated.

For the dynamical mesh refinement, we use a Jeans criterion in which
we resolve the local Jeans length with at least 8 grid cells in order
to prevent spurious fragmentation \citep{Truelove97}. Thus, the numerical grid is 
refined to reach a maximum resolution of $\Delta x
\sim 9.8 \times 10^{-4} \, \rm{pc}$ ($\sim 201 \, {\rm AU}$). Once the
maximum refinement level is reached in a given cell, no further
refinement is performed and a sink particle can be formed when the
density in this cell exceeds a threshold density, $n_{\rm thr}=1.7 \times 10^7 \, {\rm
  cm}^{-3}$. The sink particles can then
accrete mass from their surroundings (within an accretion radius of
$\sim 2.5 \Delta x$), increasing their mass. The numerical scheme for
creation of sinks and further mass accretion is that of
\citet[][]{Federrath10}. 

\begin{figure}
\includegraphics[scale=0.35,keepaspectratio=true,trim=1.3cm 0cm 0cm 1.3cm,clip]{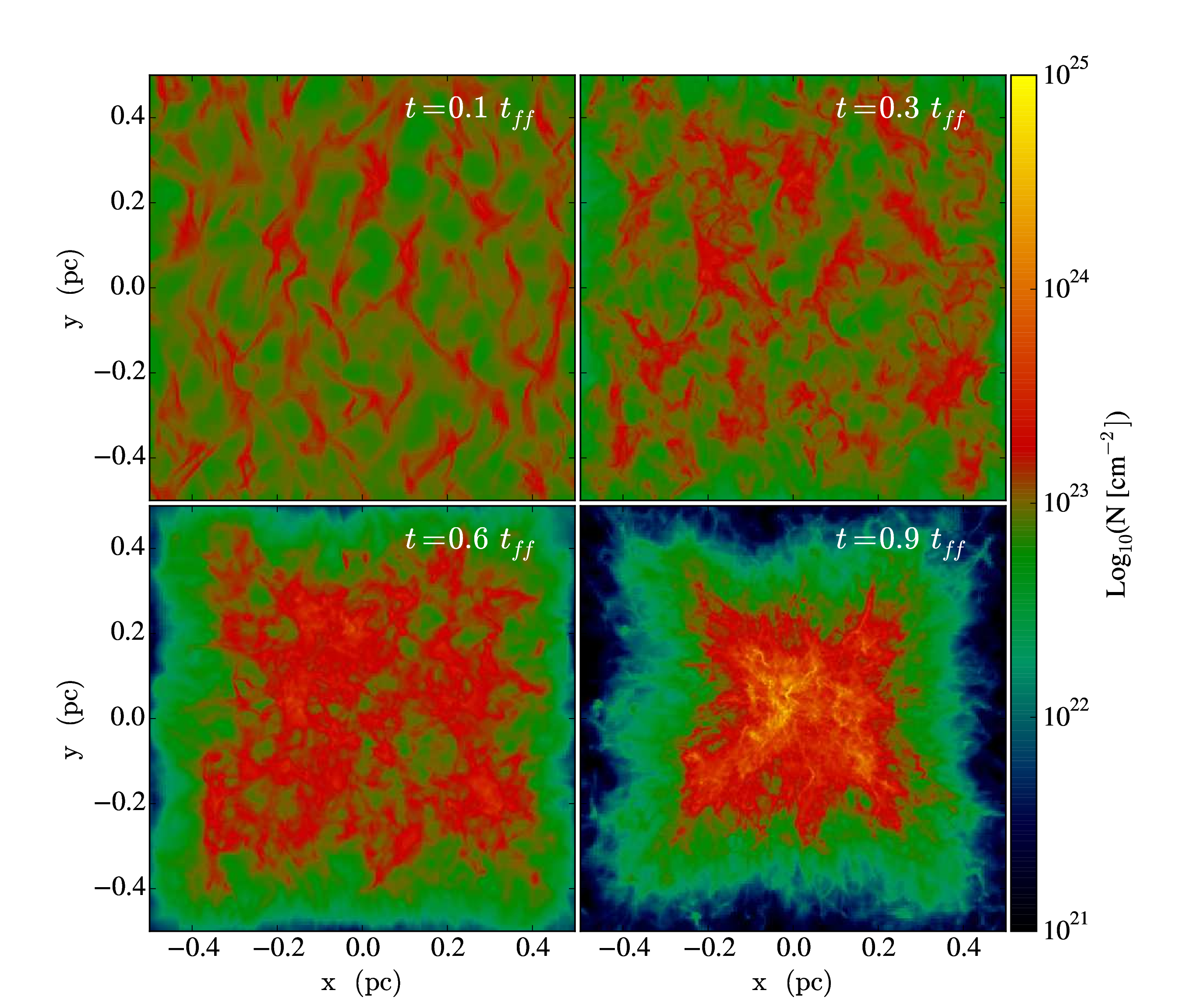}
  \caption{Column density maps at 0.1, 0.3, 0.6 and 0.9 free-fall
times ($t_{\rm ff}$). 
Note that the early evolution is dominated by turbulent motions, which dissipate energy through shocks, forming thus filamentary structures. Once the turbulence is dissipated, the gravity takes over and the gas is in free-fall dragging the magnetic field lines.
Note that the sinks particles (which start to form at 0.75~$t_{\rm ff}$) are not shown at 0.9~$t_{\rm ff}$ in order to appreciate better the complex structure of the collapsing gas.}
  \label{fig:evolution}
\end{figure}
We run the simulation until 0.9 times the free-fall time ($t_{\rm ff}$) and we
compare the numerical model with the observations at this point, because
gravity dominates the gas dynamics. In general terms, the evolution
can be considered quite similar to that shown in \citet{Ballesteros-Paredes15}: the
initial supersonic velocity fluctuations shock and rapidly dissipate
their kinetic energy, the gas then collapses in a global but
hierarchical and chaotic way forming a complex network of cores and
filaments, carrying with them the magnetic field lines. 
The time evolution of the simulation is illustrated in Fig.~\ref{fig:evolution}, which shows column density maps in the $x-y$ plane at $t=$ 0.1, 0.3, 0.6 and 0.9 $t_{\rm ff}$. As it can be seen, the first time steps are dominated by density fluctuations that grow in mass and merge as collapse proceeds. These density fluctuations are elongated mostly in the $y$-direction because the initial magnetic field is aligned along constant values of $x$, allowing the mass to flow in the $y$-direction more freely than in the perpendicular directions \citep[in contrast to the initial fluctuations of the non-magnetic simulations presented earlier in][]{Ballesteros-Paredes15}. Nevertheless, as the collapse proceeds, no preferred elongation is seen in the column density maps, and the final structure is quite similar to the non-magnetic case \citep[see Fig.~1 in][]{Ballesteros-Paredes15}, suggesting that, indeed, gravity is driving the dynamics of the dense collapsing core. 

The onset of star formation (sink formation) occurs at $0.75 \, t_{\rm ff}$, once the cloud is globally collapsing. We compare the stellar content from both the simulation and observations in Sec.~\ref{sec:comparison}.

Finally, to compare the numerical model with the simulations, we extract, at $0.9 \, t_{\rm ff}$, the central cubic subregion of 0.3~pc per side in order to match the angular size with the observations (see Fig.~\ref{fig:simulation}).

It is important to notice that even though the magnetic-to-flux ratio should be conserved in MHD simulations, this is valid only in an average sense over the whole box. However, as a consequence of the collapse the magnetic field lines increase their values locally, as in the portion of the simulation that we have analyzed (Fig.~\ref{fig:simulation}), although it starts magnetically supercritical, it ends up with equipartition values. The fact that cores and or clumps are observed in equipartition (virial) between gravity and their internal energies (magnetic, kinetic) has been interpreted as a signature of collapse: these energies do not have to know what are their detailed values to keep the cloud in equipartition, and the more reasonable physical reason for this is that gravity dominates bulk, chaotic and hierarchical motions, dragging the magnetic field as collapse proceeds \citep[e.g.,][]{Ballesteros-Paredes06,Ballesteros-Paredes11}.

\subsubsection{ARTIST}
We utilize ARTIST\footnote{URL: \url{http://youngstars.nbi.dk/artist/Welcome.html}} 
\citep[Adaptable Radiative Transfer Innovations for Submillimetre Telescopes;][]{Brinch10}
to generate a synthetic intensity cube (RA, DEC, $v_{\mathrm{LSR}}$) of the (4--3) transition of H$^{13}$CO$^+$ from the 3-D structure generated in the numerical simulation. We take a fractional abundance of $3 \times 10^{-11}$ with respect to the H$_2$ density \citep[see e.g.,][]{Girart00}, and we assume a dust temperature of 30 K, a source distance of 1.3~kpc, and no inclination/rotation, i.e., face-on view. The angular resolution is 0.05 arcsec. 

Additionally, we utilize the DustPol module of ARTIST \citep{Padovani12}, to generate a synthetic map of the dust polarized emission at $870 \, \mu\mathrm{m}$ on the plane of the sky from the density distribution of the magnetic field generated in the numerical simulation.

\subsubsection{SMA response} \label{sec:SMAresponse}
The numerical simulations and the H$^{13}$CO$^+$ (4$-$3) ARTIST spectral cube were convolved with the SMA interferometric response for a realistic comparison with the observations. This was done using the {\sc uvmodel} task in {\sc Miriad} by converting the modeled maps to visibilities using the same distribution of the visibilities in the (u, v)-plane as in the NGC 6334 V SMA observations. Thus, we filtered the same large-angular scale structures as in the real SMA observations. Later, the same parameters were used to create the final maps (e.g., pixel size,  {\sc Robust} parameter, ...). For the simulated maps, the procedure was followed independently for Stokes I, Q, and U. Once the synthetic SMA-like Stokes I, Q, U maps were obtained, the polarization intensity and position angles were obtained in the same way as for the SMA data.  

\subsection{Observations vs. simulation: polarization angular dispersion function comparison} \label{sec:SF}
From dust polarization observations we are able to estimate the magnetic field strength in the plane of the sky using the Chandrasekhar-Fermi (CF) equation \citep{CF53},
\begin{equation}
\frac{\delta B}{B}\simeq\frac{\sigma_v}{V_A},
\end{equation} 
where $B$ is the strength of the magnetic field, $\delta B$ the variation of $B$, $V_A=B/\sqrt{4\pi\rho}$ is the Alfv\'en
speed at mass density $\rho$, and $\sigma_v$ is the velocity dispersion along the line of sight. Different statistical methods have been developed in order to avoid some of the CF method caveats \citep{Hildebrand09,Houde09,Houde11,Franco10,Koch10}. 

Assuming two statistically independent components of $B$ (the large-scale magnetic field $B_0$ and the turbulent magnetic field $B_t$), \citet{Houde09} suggest that the ratio of $B_t$ to $B_0$ can be evaluated from the angular dispersion function that accounts for the polarization angle differences ($\Delta\Phi$) as a function of the distance ($l$) between the measured positions. 

The angular dispersion function can be written as
\begin{equation}
1-\langle\mathrm{cos}[\Delta\Phi(l)]\rangle\simeq\frac{\langle B_t^2\rangle}{\langle B_0^2\rangle}\frac{1}{N}\left[1-\mathrm{e}^{-l^2/2(\delta^2+2W^2)}\right]+\sum_{j=1}^{\infty} a'_{2j}~l^{2j},
\label{eq:SF}
\end{equation}
where 
\begin{equation}
N=\left[\frac{(\delta^2+2W^2)\Delta'}{\sqrt{2\pi}\delta^3}\right]
\label{eq:N}
\end{equation}
is the number of turbulent cells along the line of sight, $\delta$ is the magnetic field turbulent correlation length (assumed to be much smaller than the thickness of the cloud $\Delta'$), $W$ is the synthesized beam (i.e., FWHM/$\sqrt{8~ln(2)}$), and the summation is a Taylor expansion representing the large-scale magnetic field component which does not involve turbulence. For displacements $l$ less than a few times $W$ we keep only the first $l^2$ term in the Taylor expansion. We do not use \citet{Houde16} approximation which takes into account the interferometer filtering effect, as we want to simplify the variables to perform a qualitative comparison between the observational data and the simulation. In addition, the simulation has been convolved by the SMA response (see Section~\ref{sec:SMAresponse}), thus, the effect should be the same as in the observational data.
 
The first term on the right hand side in equation~(\ref{eq:SF}) contains the integrated turbulent magnetic field contribution, while the second (exponential) term represents the correlation by the combined effect of the beam ($W$) and the turbulent magnetic field ($\delta$). The value of the correlated component at the origin $l=0$, $f_{\mathrm{NC}}(0)$, allows us to estimate the turbulent to large-scale magnetic field strength ratio as
\begin{equation}
\frac{\langle B_t^2\rangle}{\langle B_0^2\rangle}=N~f_{\mathrm{NC}}(0).
\label{eq:Bratio}
\end{equation}
\begin{table}
\caption{Angular dispersion function fit parameters}
\begin{center}
\begin{tabular}{c c c}
\hline
\hline
Parameters									&SMA observations		&Simulations		\\
\hline
$\delta^\mathrm{a}$ (mpc)						&17, 25				&17, 25			\\
$f_{\mathrm{NC}}^\mathrm{b}$ 					&0.29, 0.52			&0.13, 0.28		\\
$\Delta'$ (pc)									&0.07				&0.07			\\
$n$ (cm$^{-3}$) 								&$6.1\times10^5$		&$6.1\times10^5$ 	\\
$\sigma_v$ (km s$^{-1}$)							&0.6					&0.6				\\
$\langle B_t ^2\rangle/\langle B_0 ^2\rangle^\mathrm{b}$	&$\simeq0.67, 0.68$		&$\simeq0.31, 0.37$	\\
$\langle B\rangle_{pos}$ (mG)						&$\simeq0.7$			&$\simeq1.1$		\\
\hline	
\end{tabular}
\end{center}
\begin{list}{}{}
\item[$^\mathrm{a}$] We used two fixed values of the parameter $\delta$.
\item[$^\mathrm{b}$] The two values correspond to the results using the two fixed values of $\delta$. 
\end{list}
\label{tab:SF}
\end{table}
\begin{figure}
\includegraphics[scale=0.31,keepaspectratio=true]{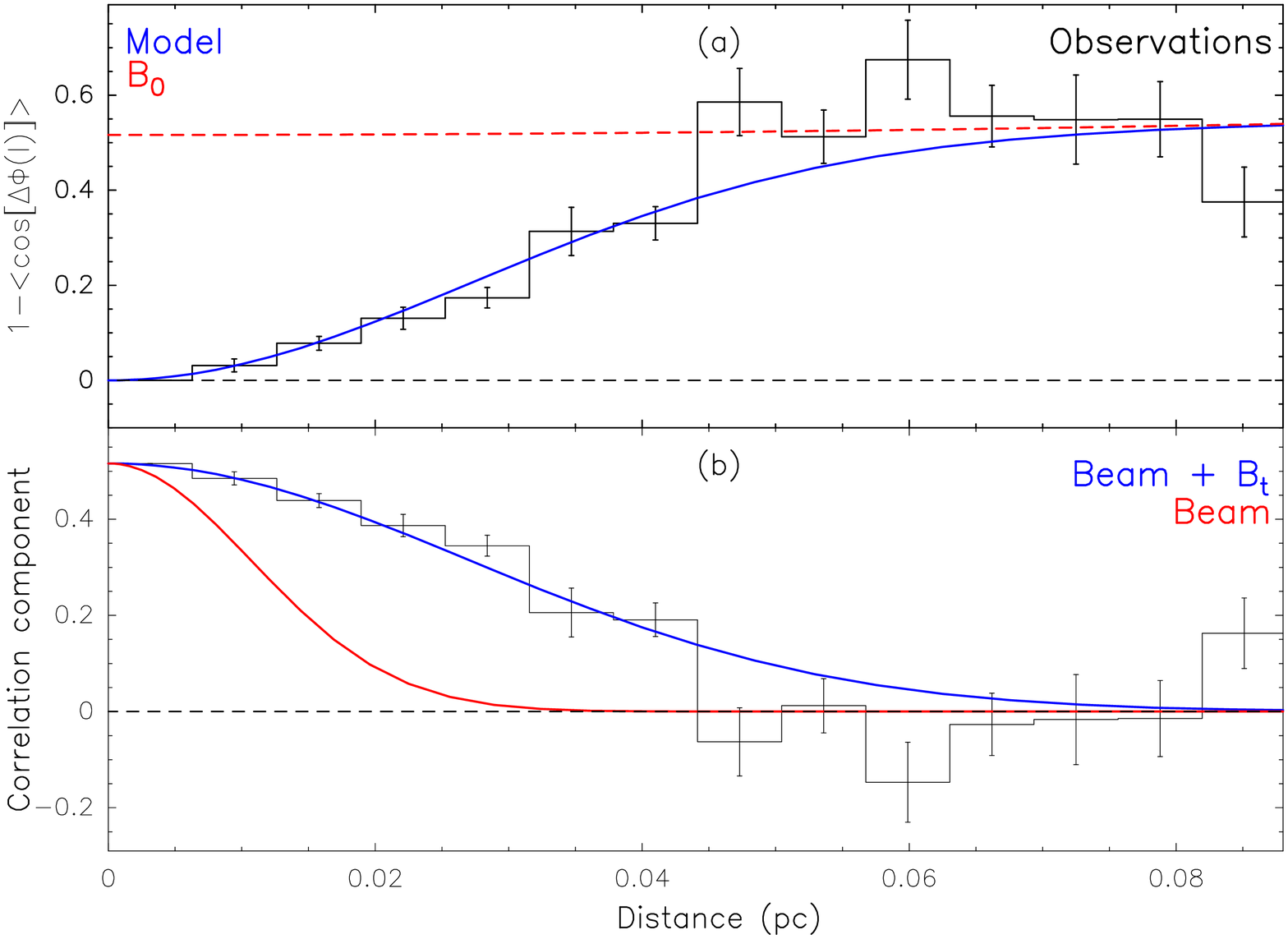}
\includegraphics[scale=0.31,keepaspectratio=true]{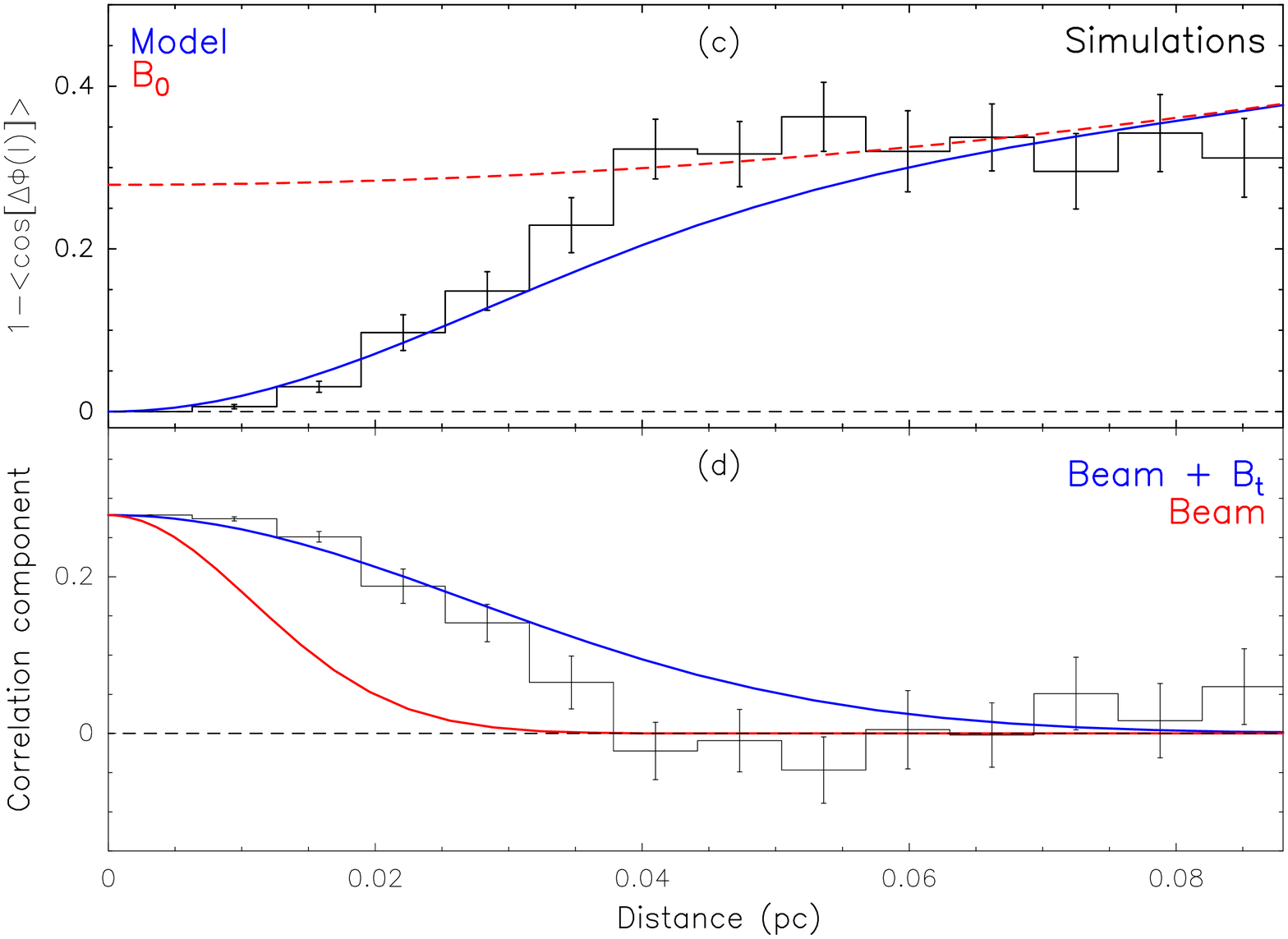}
  \caption{{\it Panels (a) and (c)}: angular dispersion function comparison from NGC~6334~V observed and simulated magnetic field segments, using a Nyquist sampling. The solid black line and error bars are the mean and standard deviation of all the pairs contained in each bin. The red dashed line does not contain the correlated part of the function (i.e., $f_\mathrm{NC}(0)+a'_{2}~l^{2}$). The blue line shows the fit to the data using eq.~(\ref{eq:SF}) and fixed $\delta=25$~mpc. {\it Panels (b) and (d)}: The solid black line represents the correlated component (exponential term of eq.~(\ref{eq:SF})) of the fit to the data. The solid red line shows the correlation due to the beam and the solid blue line shows the correlation due to the beam and the turbulent component of the magnetic field. Panels (a) and (b) show the results from the SMA observations. Panels (c) and (d) show the results from the numerical simulations.}
  \label{fig:SF}
\end{figure}

To fit the angular dispersion function to our results and to be able to perform a qualitative comparison between the observations and the numerical simulations, we used two fixed values of the parameter $\delta$, 17~mpc \citep[from][]{Girart13} and 25~mpc which produced a better fit (see Fig.~\ref{fig:SF}). The derived value of the correlated component at the origin is $f_\mathrm{NC}\simeq0.29, 0.52$ and $\simeq0.13, 0.28$ for both values of $\delta$ and the observations and simulations respectively (see Table~\ref{tab:SF}). A reasonable approximation to the core's effective thickness, $\Delta'$, is to assume that it is similar to the average diameter of the dense core measured in the plane of the sky with the SMA \citep{Koch10}, which in our case is $\simeq0.07$~pc (see Section~\ref{continuum}). Using eqs.~(\ref{eq:N}) and (\ref{eq:Bratio}) $\langle B_t ^2\rangle/\langle B_0 ^2\rangle$ is $\simeq0.7$ for the observations, i.e., it is close to equipartition between the turbulent and ordered magnetic field energies. This is similar to the results obtained towards the high-mass star-forming region DR21(OH) \citep{Girart13}. However, for the simulations the values are lower ($\simeq0.3, 0.4$). One possibility for this difference is that the size of the core in the simulations could be larger than the observed one, as $N$ is proportional to $\Delta'$. Also, the thermal Q and U instrumental noise produced by the SMA can induce an increase of the angular dispersion which is not taken into account in the simulations (see upper panel of Fig.~\ref{fig:SF}). 

To estimate the large-scale magnetic field strength in the plane of the sky we used the CF equation, $\langle B_0^2\rangle^{1/2}=\sqrt{4\pi\rho}~\sigma_v~[\langle B _t ^2\rangle/	\langle B _0 ^2\rangle]^{-1/2}$, \citep[e.g., see Equation (57) by][]{Houde09}. 
Table~\ref{tab:SF} shows the values used for the velocity dispersion $\sigma_v$ and the number density $n$ (using a mean molecular weight of 2.33). To derive the averaged volume density we used the mass derived from the total 870 $\mu$m flux density detected with the SMA ($\sim50$~M$_{\odot}$) and the size of the core ($\sim0.07$~pc). We estimated the velocity dispersion from the H$^{13}$CO$^+$ (4--3) data, because it is well correlated with the 870~$\mu$m dust emission (see upper panel of Fig.~\ref{fig:mom0}). The derived value for the ordered large-scale magnetic field strength in the plane of the sky for both observations and simulations is $\langle B_0^2\rangle^{1/2}\simeq1$~mG. This value is in very good agreement with the average value of the magnetic field (weighted by density) measured in the simulation at 0.9~$t_{\rm ff}$, 0.9~mG.
\section{Discussion}
\label{sec:discussion}
\subsection{Simulation results and comparison with observations} \label{sec:comparison}
\begin{figure}
 \centering
\includegraphics[scale=0.45,keepaspectratio=true]{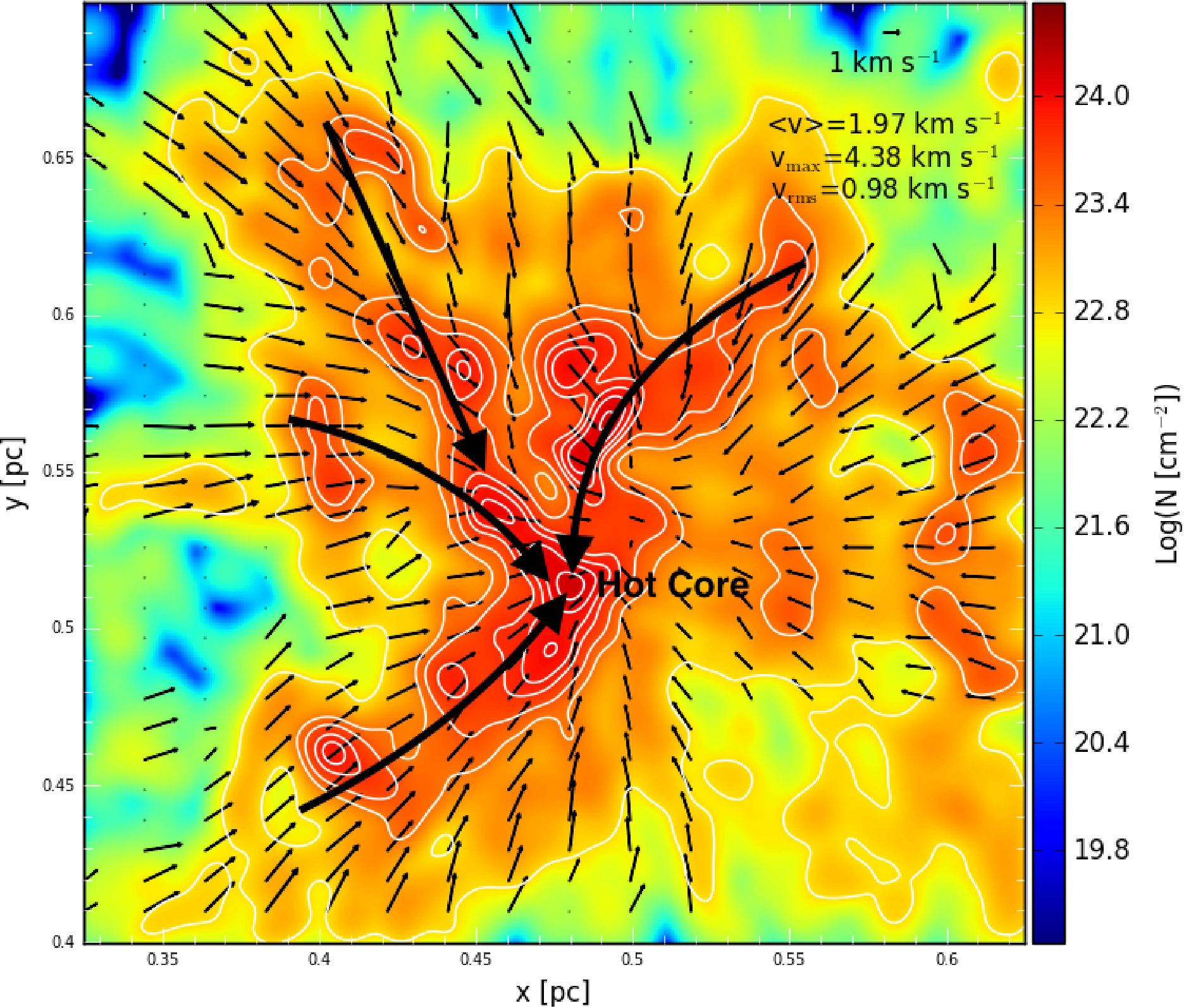}
  \caption{Column density map of the central 0.3~pc subregion of the numerical simulation at $0.9 \, t_{\rm ff}$ (see the lower right panel of Fig.~\ref{fig:evolution}). Note that the column density is convolved at the resolution of the observations. The thin black arrows represent the integrated velocity field (weighted by density) along the line of sight ($z$-direction), whereas the thick black arrows highlight the filamentary structures converging at the densest ``hot core'' region.
}
  \label{fig:simulation}
\end{figure}
In the bottom panel of Figure~\ref{fig:modelpanels} we present the results from the simulations filtered out with the SMA response. The map shows different velocity structures at scales of $\sim0.1$ pc which seem to converge towards the strongest peak of the continuum and with a similar velocity range as the one present in the observed data. As in the case of the observations, in this figure we notice that the field has three main orientations: segments that are almost horizontal, but that have two components: at +70 and +100 degrees, (also seen as a double peak in the histogram of the lower panel of Fig.~\ref{distPA}). The third component is not highly populated, but it is clearly traced by the almost vertical white segments shown in Fig.~\ref{fig:modelpanels}. This component corresponds to the long tail towards small angles of the 70-degree component in the histogram shown in the lower panel of Fig.~\ref{distPA}. Furthermore, note that at least the two main components, at +70 and +100 degrees, coincide  with the two different velocity structures traced by H$^{13}$CO$^+$.

The magnetic field configuration shown in the lower panel of Fig.~\ref{fig:modelpanels} can be understood in terms of the dynamics of the core. In Fig.~\ref{fig:simulation} we show the gas column density (colors) and the $x-y$ velocity field of the gas (thin black arrows) after evolving the simulation. Note that  this representation has no observational counterpart, and although the simulation resolution is $\sim200$ AU, the column density map shown in this figure has been convolved with a beam of the same size of the observations. The column density map shows filamentary structures which seem to converge at the center of the box, where the highest column densities are found. The velocity field shows how the gas is converging from the dense structures at the large scale ($\sim$0.2 pc) towards the highest density region, at $\sim$0.02 pc scales, and they appear to be almost radial. Thus, it is clear that the core is collapsing. The originally horizontal magnetic field has been advected at this time by the collapsing gas, resulting in an hourglass shape with a dominant nearly horizontal component at the center of the core, with position angles of $+90\pm 20$ degrees. This dominance is just the consequence of the B-field being compressed and dragged by the collapsing gas. We notice also that in the outer parts of the core, the B-field does have components that are far from parallel to the original magnetic field. Although it is a coincidence that the simulations exhibit three main orientations of the magnetic field, as in the observations, the important point here is that the B-field, in the innermost denser part of the core, seems to be oriented mostly horizontal, while in the outermost parts it looks more radial.  

It can be asked whether the selected angle of view of the simulations affects what we observe in Fig.~\ref{fig:modelpanels}. Certainly, as the magnetic field is a vector quantity, any point of view could be considered as ``particular''. And we have chosen a point of view that is somehow special: it is perpendicular to the initial magnetic field (the box is seen along the $z$-axis, while the initial magnetic field runs along the $x$-axis). Given this configuration, and since the collapse occurs almost radially, one may assume that the density and magnetic field must have, in a statistical sense, an axial symmetry with respect to a declination offset=0 (Fig.~\ref{fig:modelpanels}). This means that, in principle, if we had observed the core along any other line of sight that was perpendicular to the $x$-axis, we would have observed something similar: nearly horizontal orientations of the field in the innermost parts of the core, and more radially directed orientations in the external parts of the core. However, if we were observing the simulation along the $x$-axis, the horizontal, dominant component of the B-field would almost not be observable. The B-field should look almost radial in the outskirts of the core, with a decreasing component at the center. The more general situation will be something in between these two possibilities, i.e., radial lines in the external parts of the core, and a less-dominant magnetic field in the center of the box.

As we do not include feedback processes in the numerical model, we cannot model the hot core itself (and the  associated outflows). However, in the densest part of the simulation, where we locate the ``Hot Core'' labeled in Figure~\ref{fig:simulation}, we have a group of three stars (sinks) with a total mass of $\sim9.7$ M$_{\odot}$ in a compact space of 0.03 pc ($\sim4.7''$, assuming a distance of 1.3 kpc), which is comparable with the synthesized beam of our observations. It is remarkable that this mass is comparable with the mass inferred from kinematics arguments in the observed hot core ($\sim6$ M$_{\odot}$; see Section~\ref{velocity}), which give us confidence that we are comparing our observations with an adequate model.

Finally, in addition to the good qualitative comparison of the velocity and magnetic field structures between the SMA observations and the numerical simulations, the angular dispersion function analysis also reveals a very similar behavior for both (see Section~\ref{sec:SF} and Fig.~\ref{fig:SF}). The simulations, thus, again seem to be a good approximation to the observational results.

\subsection{Outflow-generated cavity}
We have seen in Section~\ref{sec:molecules} that the emission from the extended dense core of NGC 6334 V (traced by H$^{13}$CO$^+$\,(4--3) and CH$_3$OH 7(1,7)$-$6(1,6)) presents a clear ring-like structure around the systemic velocity, $-5.7$ km s$^{-1}$ (see Fig.~\ref{fig:mom0}). The infrared and radio sources found in the region are located at the center of the ring, that also coincides with the cavity of the dust continuum emission. In addition, we detected high-velocity gas traced by the CO (3$-$2) molecular line at the position of the cavity and at the center of the ring structure (see Fig.~\ref{COpanels}). 4.5 $\mu$m emission is also detected crossing through the center of the ring with an east-west elongated shape presenting a bow shock structure towards the west, suggesting the emission could be shock- or outflow-generated (see Fig.~\ref{spitzer}). These results suggest the presence of an east-west outflow \citep[also reported by][]{Hashimoto07,Simpson09} that could be affecting the central region of NGC 6334 V, causing a cavity in both the extended dense gas as well as in the dust continuum structure. 

\subsection{Comparison with other NGC 6334 star-forming sites}
In a previous work, \citet{Li15} studied the magnetic field properties in several star-forming regions inside NGC 6334,  from 100 down to 0.01 pc scales. Source V was not included in this study as large-scale polarization observations were not available. They concluded that the well ordered magnetic field (at all scales) in sources I, I(N) and IV plays a crucial role in the fragmentation of NGC 6334. 

However, in NGC 6334 V, we see a more complex magnetic field morphology governed by the gas dynamics. It seems that NGC 6334 V, which is located at the southern end of the NGC 6334 ridge and possibly at the intersection region with another northwest-southeast gas filament, with gravity already being the dominant force over the magnetic field \citep[e.g.,][]{Andre16}, may be more dynamically evolved than the sources I, I(N) and IV \citep[note that the total mass in NGC 6334 V seems to be smaller than in source I and I(N);][]{Hunter06}. 

\section{Conclusions}
\label{sec:conclusion}
We analyzed SMA observations at 345 GHz towards the intermediate/high-mass star-forming region NGC~6334~V. The main results are as follows:
\begin{itemize}
\item The dust continuum emission presents an arc-like structure of $\sim$0.07 pc ($\sim$14,000 AU), with three distinguishable peaks forming a cavity structure (Fig.~\ref{fig:continuum}). The total mass of the arc-like molecular structure is $\sim50$~M$_{\odot}$. The main peak (HC) has a mass of $\sim4$--9~M$_{\odot}$ and exhibits the typical chemistry of a hot core with emission from $^{34}$SO, SO$_2$, $^{34}$SO$_2$, CH$_3$OH, CH$_3$OCH$_3$, CH$_3$OCHO and HC$_3$N. Dense core tracers H$^{13}$CO$^+$ (4--3) and CH$_3$OH 7(1,7)$-$6(1,6) show extended emission forming a ring-like structure at the systemic velocity of $-5.7$ km s$^{-1}$. The presence of an outflow at the center of the region could be affecting the surrounding gas forming the cavity traced by the dust continuum and the dense gas emission from H$^{13}$CO$^+$ and CH$_3$OH.
\item Kinematically, the dense core has two distinctive velocity components, one at $-5.0$ km s$^{-1}$ tracing a filamentary-like arm on the southeast part of the core and another one at $-7.4$ km s$^{-1}$ arising mainly from the northern part (Figs.~\ref{fig:mom1},~\ref{fig:pvcuts}). Both velocity structures converge towards the hot core. At higher density regions in the hot core, the hot core line tracers CH$_3$OCHO 31(0,31)$-$30(1,30) and several transitions of CH$_3$OH, CH$_3$OCH$_3$, $^{34}$SO$_2$ and SO$_2$ show the same two velocity components. The shock tracer SiO (8--7) (Fig.~\ref{fig:pvcuts}b,c) presents a velocity component just at an intermediate velocity ($\sim-6$ km s$^{-1}$) suggesting interaction between the two flows.
\item The magnetic field (derived from the dust thermal polarized emission at 870 $\mu$m) shows a bimodal converging pattern towards the hot core and follows the distribution of the two velocity components.
\item We produced synthetic observations from numerical simulations of massive star-forming regions dominated by gravity. As in the SMA observations, the numerical simulations produced two distinctive velocity components (traced by ARTIST-generated H$^{13}$CO$^+$ emission) converging towards the strongest peak of the dust thermal emission, with the magnetic field being dragged by the gas. The polarization angular dispersion function comparison between the observations and simulations also reveals a very similar behavior.
\item  NGC 6334 V may be more dynamically evolved than the sources I, I(N) and IV \citep[see][]{Li15}, with gravity already being the dominant force over the magnetic field. 
\end{itemize}
Finally, these results show how the gas is being accreted from the larger scale extended dense core ($\sim0.1$ pc) of NGC 6334 V towards the higher density hot core region at $\sim0.02$ pc scales, through two distinctive converging flows dragging along the magnetic field whose strength seems to have been overcome by gravity.
\acknowledgments
\begin{small}
We would like to thank the referee for her/his useful comments. We also thank the SMA staff for their support which makes these studies possible. The Submillimeter Array is a joint project between the Smithsonian Astrophysical Observatory and the Academia Sinica Institute of Astronomy and Astrophysics, and is funded by the Smithsonian Institution and the Academia Sinica. We thank Mauricio Tapia for thoughtful discussions about the origin of the infrared emission in NGC 6334 V. We thank Marco Padovani for his valuable support with ARTIST, and Gilberto Zavala P{\'e}rez and Alfonso H. Ginori Gonz{\'a}lez for their computational support. CJ acknowledges support from MINECO (Spain) BES-2012-052481 grant. CJ and JMG acknowledge support from MICINN (Spain) AYA2014-57369-C3 grant. JMG also acknowledges the support from the MECD (Spain) PRX15/00435 travel grant. AP and MZA acknowledge financial support from UNAM-DGAPA-PAPIIT IA 102815 grant (Mexico). 
MZA also acknowledges CONACyT for a postdoctoral fellowship at the University of Michigan. PMK acknowledges support from the Ministry of Science and Technology in Taiwan through grant MoST 103-2119-M-001-009 and an Academia Sinica Career Development Award. J.B.P. acknowledges UNAM-PAPIIT grant number IN110816, and to UNAM's DGAPA-PASPA Sabbatical program. He also is indebted to the Alexander von Humboldt Stiftung for its invaluable support.\\
\end{small}

\bibliographystyle{apj} 
\bibliography{bibliography} 



\end{document}